\documentstyle[pre,aps,epsf,multicol,tabularx]{revtex}
\tighten
\begin{document} 
   
\newcommand{\bfGamma}{\mbox{\boldmath $\bf\Gamma$}}
\newcommand{\smallbfGamma}{\mbox{\boldmath $\scriptstyle \Gamma$}}
\newcommand{\caGamma}{\mbox{$\it\Gamma$}}
\newcommand{\caA}{{\cal{A}}}
\newcommand{\bfq}{\mbox{\boldmath $\bf q$}}
\newcommand{\bfp}{\mbox{\boldmath $\bf p$}}
\newcommand{\bfx}{\mbox{\boldmath $\bf x$}}
\newcommand{\bfXi}{\mbox{\boldmath $\bf \Xi$}}

\newcommand{\lines}{\baselineskip 0.386cm}
\newcommand{\linesabs}{\baselineskip 0.35cm}
\lines
\renewcommand{\baselinestretch}{0.85}


\draft

\title{Stepwise structure of Lyapunov spectra for many particle systems 
by a random matrix dynamics}

\author{Tooru Taniguchi and Gary P. Morriss}

\address{School of Physics, University of New South Wales, Sydney, 
New South Wales 2052, Australia}

\date{\today}

\maketitle

\vspace{0cm}
\begin{abstract}

\linesabs

   The structure of Lyapunov spectra for many particle systems 
with a random interaction between the particles is discussed.
   The dynamics of the tangent space is expressed as a master 
equation, which leads to a formula that connects the positive 
Lyapunov exponents and the time correlations of the particle interaction 
matrix. 
   Applying this formula to one and two dimensional models 
we investigate the stepwise structure of the Lyapunov spectra, 
which appear in the region of small positive Lyapunov exponents. 
   Long range interactions lead to a clear separation of the Lyapunov 
spectra into a part exhibiting stepwise structure and a part changing 
smoothly. 
   The part of the Lyapunov spectrum containing the stepwise structure 
is clearly distinguished by a wave like structure in the eigenstates of 
the particle interaction matrix.
   The two dimensional model has the same step widths as found 
numerically in a deterministic chaotic system of many hard disks. 
\end{abstract}

\vspace{0.1cm}

%
%
%
%
%
%
%
 
\vspace{0.3cm}

\begin{multicols}{2}

\narrowtext
  

\section{Introduction}

   Chaos is characterized by a rapid expansion of a small initial error, 
and the Lyapunov exponent, which is defined as the time averaged exponential 
rate of the time evolution of infinitesimal perturbations of the dynamical 
variables, is introduced to express such characteristics quantitatively. 
   The Lyapunov exponent is used to discuss some statistical properties 
of many body systems (eg. the mixing property). 
   It is also known that the Lyapunov exponents are connected to the amount 
of information of the system (eg. the Kolmogorov-Sinai entropy), 
and to transport coefficients (eg. conductance and viscosity) 
\cite{Eva90a,Gas98,Dor99}. 

   In general, the Lyapunov exponent depends on the direction of the infinitesimal 
perturbation of the dynamical variables at an initial time, so we obtain 
a Lyapunov exponent for each independent direction in the system. 
   The sorted set of such Lyapunov exponents, is the so called 
Lyapunov spectrum, and has been the subject of study in many 
chaotic particle systems. 
   For example, the thermodynamic limit of the Lyapunov spectrum, 
that is, that the spectrum retains its shape as the number of 
particles increases, is discussed using numerical evidence 
\cite{Liv86,Liv87}, random matrix approaches \cite{Pal86,New86,Eck88}, 
a periodic orbit approach \cite{Tan01} and mathematical arguments  
\cite{Rue82,Sin96}. 
   Some works showed a linear behavior of the Lyapunov 
spectrum in one dimensional models with nearest neighbor 
interactions \cite{Liv86,Liv87,Pal86,New86,Eck88}, although 
this conjecture is modified in weak chaotic systems \cite{Yam98,San00}. 
   The effect of the rotational degrees of freedom of molecules on the 
Lyapunov spectrum was investigated in a model consisting of diatomic 
molecules, which showed an explicit separation 
of the rotational and translational degrees of freedom if the departure 
from sphericity is small enough\cite{Mil98a,Mil98b}.
   The Lyapunov spectra  of systems in non-equilibrium steady states were 
investigated by Refs. \cite{Pos88,Mor88,Mor89,Del96} with the 
discovery of the conjugate pairing rule for some systems with 
isokinetic thermostats \cite{ECM1,Sar92,Det96,Mor02}.
   Avoided crossings and level repulsion in the Lyapunov spectrum, similar 
to the behavior of the energy levels in quantum chaotic system, was 
discussed in a mutually coupled map system \cite{Ahl01}.  
   These works suggest the possibility of getting information about the system 
from the structure of its Lyapunov spectrum.
   The Lyapunov spectrum has also been of interest for the 
number dependence of the maximum Lyapunov exponent in particle systems
\cite{Pal86,Del96,Sea97,Zon98}, 
and the Kaplan-York Lyapunov dimension \cite{Gra87,Hoo94,Wag00,Eva00}. 
   
   One of the characteristics of the Lyapunov spectrum, which has been shown  
recently in systems consisting of many hard disks, is its stepwise 
structure \cite{Mil98a,Mil98b,Del96,Pos00,Mcn01a}. 
   Such a stepwise structure appears in the distinct region of small positive 
Lyapunov exponents. 
   Corresponding to this stepwise structure are the so called Lyapunov modes, 
a wave like structure in the tangent space of each eigenvector of a 
degenerate Lyapunov exponent, that is, for each stepwise structure. 
   An explanation of the stepwise structure of the 
Lyapunov spectrum due to symmetries of the system was tried using 
two dimensional periodic orbit models \cite{Tan01}.
   On the other hand, explanations for the Lyapunov modes were 
suggested using a random matrix 
approach for a one dimensional model \cite{Eck00} and 
more recently using a kinetic theoretical approach \cite{Mcn01b}.
   The small Lyapunov exponents correspond to slow relaxation processes   
in the equilibrium state, so these results suggest the possibility of
characterizing the
macroscopic behavior in many particle systems from a microscopic point of view. 

   This paper has two main purposes. 
   First we consider the dynamics of the tangent space in  
dynamical systems with a random interaction between the particles.
   Different from the other random matrix approaches for the 
Lyapunov spectrum \cite{Pal86,Eck88,Eck00} 
which use discretized models in time, 
we consider the continuous dynamics in time and derive 
a method to calculate the Lyapunov spectrum from a master equation technique. 
   Using this method the Lyapunov spectrum is directly connected to the 
correlations of the particle interaction matrix. 
   As the second purpose we discuss the stepwise structure of the Lyapunov 
spectrum using this random matrix approach. 
   We consider one and two dimensional models satisfying  
the total momentum conservation with periodic boundary condition.
   These models show a stepwise 
structure of the Lyapunov spectra only in a region of small positive 
Lyapunov exponents.    
   We discuss the robustness of the stepwise structure against perturbations 
in the matrix elements of the particle interaction matrix. 
   The long range interactions between the particles lead to a clear separation 
into a part of the spectrum exhibiting stepwise structure and a part changing 
smoothly with exponent number. 
   We also find a wave like structure in the eigenstates of the particle 
interaction matrix in that part of Lyapunov spectrum containing stepwise 
structure.
   As a high dimensional effect we show that in the two dimensional model 
we get wider steps in the Lyapunov spectrum, such as steps 
consisting of 4 and 8 degenerate exponents, rather than simply 2 
exponents as in the one dimensional models. 
   The step widths of 4 and 8 exponents are actually observed in 
the numerical simulation of the many hard-core disk system \cite{Pos00}.


\section{Random Matrix Dynamics of the Tangent Vector}

   We consider a Hamiltonian system whose state at time $t$ is described 
by the $2N$ dimensional phase space vector $\bfGamma (t) \equiv  
(\bfq(t),\bfp(t))^{T}$ with the coordinate vector $\bfq(t)$, the momentum 
vector $\bfp(t)$ and the transpose operation $T$.  
   The dynamics of the phase space is described by the Hamilton equation 

\begin{eqnarray}
   \frac{d\bfGamma (t)}{dt} 
   = J \frac{\partial H(\bfGamma (t),t)}{\partial 
   \bfGamma (t)}
\label{HamilEquat}\end{eqnarray}

\noindent with the Hamiltonian $H(\bfGamma (t), t)$.  
   Here we allow an explicit time-dependence of the Hamiltonian, and 
$J$ is the $2N \times 2N$ matrix defined by 
   
\begin{eqnarray}
   J \equiv \left(\begin{array}{cc} 
   \underline{0} & I \\
   -I & \underline{0}
   \end{array}\right)
\label{MatriJ}\end{eqnarray}

\noindent where $I$ is the $N \times N$ identical matrix and $\underline{0}$ 
is the $N \times N$ null matrix.
   It should be noted that the matrix $J$ satisfies the conditions 
$J^{T}=-J$ and $J^{2}=-\bar{I}$ with the $2N \times 2N$ identical matrix 
$\bar{I}$. 

   The dynamics of the tangent vector $\delta\bfGamma (t)$ as a small 
deviation of the phase space vector is described by the linearized equation 

\begin{eqnarray}
   \frac{d\delta\bfGamma (t)}{dt} 
   = J L(t) \; \delta\bfGamma (t)
\label{TangeEquat}\end{eqnarray}

\noindent of the Hamiltonian equation (\ref{HamilEquat}) 
where $L(t)$ is defined by 

\begin{eqnarray}
   L(t) \equiv 
   \frac{\partial^{2} H(\bfGamma (t), t)}{\partial \bfGamma (t) 
   \partial\bfGamma (t)}.
\label{MatriL}\end{eqnarray}

\noindent The matrix $L(t)$ is symmetric: $L(t)=L(t)^{T}$. At this 
point we note that an equation equivalent to Eq. (\ref{TangeEquat}) can also be 
obtained for system whose dynamics is not derivable from a Hamiltonian.

   Equation (\ref{TangeEquat}) is formally solved as  

\begin{eqnarray}
   \delta\bfGamma (t) 
   = M(t,t_{0}) \; \delta\bfGamma (t_{0})
\label{TangeEquat2}\end{eqnarray}

\noindent where the matrix $M(t,t_{0})$ is defined by

\begin{eqnarray}
   M(t,t_{0}) \equiv \stackrel{\leftarrow}{T} 
   \exp{\left[J \int_{t_{0}}^{t}ds\;L(s)\right]}
\label{MatriM}\end{eqnarray}

\noindent
with the positive time-ordering operator 
$\stackrel{\leftarrow}{T}$ with the latest time to the left.
Equation (\ref{MatriM}) leads to 

\begin{eqnarray}
   M(t,t_{0})^{-1} &=& \stackrel{\rightarrow}{T} 
   \exp{\left[-J \int_{t_{0}}^{t}ds\;L(s)\right]} 
   \nonumber \\ 
   &=& -J M(t,t_{0})^{T} J ,  
\label{MatriMPrope}\end{eqnarray}

\noindent  with the negative time-ordering operator 
$\stackrel{\rightarrow}{T}$ with the latest time to the right,  
so that the Hamiltonian phase volume is 
preserved, namely $|\mbox{Det}\{ M(t,t_{0})\}|=1$. 

   For simplicity, we consider the case of zero magnetic field, so 
the Hamiltonian $H(\bfGamma (t), t)$ is represented as 
$H(\bfGamma (t),t) = |\bfp(t)|^{2}/(2m)+V(\bfq(t),t)$, where 
$m$ is the mass of particle and $V(\bfq(t),t)$ is the potential 
energy at time  $t$. 
   In this case the matrix $L(t)$ is given by 

\begin{eqnarray}
   L(t) = \left(\begin{array}{cc}
      - R(t)          & \underline{0} \\
      \underline{0}   & I/m
   \end{array} \right)
\label{Hamil}\end{eqnarray}

\noindent where $R(t)\equiv(R_{\alpha\beta}(t))$ is  the 
$N \times N$ symmetric matrix defined by 
$-\partial^{2}V(\bfq(t), t)/\partial\bfq(t)\partial\bfq(t)$. 
   The effects of particle interactions and fixed scatterers 
are taken into account through the matrix $R(t)$. 
   
  Now we consider the case where the particle interactions or collisions with 
fixed scatterers occur randomly enough, so that the matrix elements 
$R_{\alpha\beta}(t)$, $\alpha=1,2,\cdots,N$, $\beta=1,2,\cdots,N$ 
can be regarded as Gaussian white randomness in the sense of     

\begin{eqnarray}
   \langle R_{\alpha_{1}\beta_{1}}(t_{1}) 
      R_{\alpha_{2}\beta_{2}}(t_{2})
      \cdots
      R_{\alpha_{2n-1}\beta_{2n-1}}(t_{2n-1})\rangle = 0 , 
\label{WhiteNoise1}\end{eqnarray}
\begin{eqnarray}   
   && \langle R_{\alpha_{1}\beta_{1}}(t_{1}) 
       R_{\alpha_{2}\beta_{2}}(t_{2}) 
      \cdots
      R_{\alpha_{2n}\beta_{2n}}(t_{2n})
      \rangle \nonumber \\
   && \hspace{0.6cm} =
      \sum_{P_{d}} 
      D_{\alpha_{j_{1}}\beta_{j_{1}}\alpha_{j_{2}}\beta_{j_{2}}} 
      D_{\alpha_{j_{3}}\beta_{j_{3}}\alpha_{j_{4}}\beta_{j_{4}}} 
     \nonumber \\
	 && \hspace{1cm}  \cdots     
D_{\alpha_{j_{2n-1}}\beta_{j_{2n-1}}\alpha_{j_{2n}}\beta_{j_{2n}}} 
      \nonumber \\
   && \hspace{1cm} \times  
      \delta(t_{j_{1}}-t_{j_{2}})
      \delta(t_{j_{3}}-t_{j_{4}})
      \cdots
      \delta(t_{j_{2n-1}}-t_{j_{2n}})
\label{WhiteNoise2}\end{eqnarray}

\noindent for any integer $n$, where we take the sum over only the 
permutation $P_{d}:$ $ (1,2,\cdots$ $,2n)$ $\rightarrow$ $ 
(j_{1}, j_{2}, \cdots,j_{2n})$, and the bracket $\langle\cdots\rangle$ 
means the ensemble average over random processes.
   Here $D_{jkln}$ is a $4$-th rank tensor and is assumed to be 
constant in the rest of this paper. 
   The tensor $D_{jkln}$ satisfies the condition
 
\begin{eqnarray}
   D_{lnjk} = D_{jkln},  
\label{DiffuConst1}\end{eqnarray}

\noindent because the equation 
$D_{jkln} 
\delta(s-t) $ $=\langle 
R_{jk}(s) R_{ln}(t) 
\rangle $ $= \langle 
 R_{ln}(t)R_{jk}(s)
\rangle $ $= D_{lnjk} 
\delta(s-t)$ is satisfied at arbitrary times $s$ and $t$.
  The symmetric property of the matrix $R(t)$ also imposes the conditions 

\begin{eqnarray}
   D_{jknl} 
   = D_{kjln}
   = D_{jkln}
\label{DiffuConst2}\end{eqnarray}

\noindent for the tensor $D_{jkln}$. 

   Under the Gaussian white conditions (\ref{WhiteNoise1}) and 
(\ref{WhiteNoise2}) for the matrix $R(t)$, the dynamics  
described by Eq. (\ref{TangeEquat}) for the tangent vector can 
be regarded as a stochastic process. 
   Now we consider the description of this stochastic process by 
a master equation for 
the probability density $\rho(\delta\bfGamma,t)$ for the tangent 
vector space in which $\rho(\delta\bfGamma,t)d\delta\bfGamma$ is 
introduced as the probability of finding a tangent vector $\delta\bfGamma$ 
in the region $(\delta\bfGamma, \delta\bfGamma+d\delta\bfGamma)$.
   By applying the Kramers-Moyal expansion to the dynamics 
(\ref{TangeEquat})  with the randomness given by Eqs. 
(\ref{WhiteNoise1}) and (\ref{WhiteNoise2}) we obtain

\begin{eqnarray}
   &&\frac{\partial \rho(\delta\bfGamma,t)}{\partial t} 
    \nonumber \\
	&& \hspace{0.5cm} 
	=  -\sum_{\alpha=1}^{N} \frac{\delta p_{\alpha}}{m} \frac{\partial 
   \rho(\delta\bfGamma,t)}{\partial \delta q_{\alpha}} 
   \nonumber \\
   &&\hspace{0.85cm} + \sum_{\alpha=1}^{N}\sum_{\beta=1}^{N} 
   \sum_{\mu=1}^{N}\sum_{\nu=1}^{N}
   \frac{1}{2} D_{\alpha\mu \beta\nu} \delta q_{\alpha} 
   \delta q_{\beta} \frac{\partial^{2} \rho(\delta\bfGamma,t)}{\partial 
   \delta p_{\mu} \partial \delta p_{\nu}}.
\label{Fokke}\end{eqnarray}

\noindent  Equation (\ref{Fokke}) is a master equation for the tangent vector  
$\delta\bfGamma = (\delta\bfq, \delta\bfp)^{T}$ with 
$\delta\bfq\equiv(\delta q_{1},\delta q_{2},\cdots,\delta q_{N})^{T}$ and 
$\delta\bfp\equiv(\delta p_{1},\delta p_{2},\cdots,\delta p_{N})^{T}$, 
especially the type of the equation called Fokker-Plank equation 
because its right-hand side includes up to the second derivative of 
the probability density with respect to the variables.
   The derivation of Eq. (\ref{Fokke}) is given in Appendix \ref{Maste}.
   We should notice that the first term of the right-hand side of Eq. 
(\ref{Fokke}) has the same form as the corresponding term in the 
Liouville equation describing non-interacting particle systems, 
because the dynamics of the phase space point and the tangent space 
point coincide in the non-interacting particle system. 
   The characteristics of the system are introduced through the 4-th rank
tensor $D_{\alpha\mu \beta\nu} $ and a boundary 
condition for a solution of Eq. (\ref{Fokke}).


\section{Lyapunov Spectrum in the Random Matrix Approach}
\label{Lya} 

   In this section, using the Fokker-Plank equation (\ref{Fokke}) for the tangent 
vector we investigate the absolute values of the tangent 
vectors, whose asymptotic behaviors lead to the Lyapunov exponents. 

   We introduce the quantities 
$\Upsilon_{jk}^{(l)}(t)$, $l=1,2,3$ as  

\begin{eqnarray}
   \Upsilon_{jk}^{(1)}(t) &\equiv& \left\langle \delta q_{j} 
      \delta q_{k} \right\rangle_{t} \label{Upsil1}\\
   \Upsilon_{jk}^{(2)}(t) &\equiv& \left\langle 
   \left(\delta q_{j}
      \delta p_{k} + \delta q_{k} 
      \delta p_{j } \right)\right\rangle_{t} /2\label{Upsil2}\\
   \Upsilon_{jk}^{(3)}(t) &\equiv& \left\langle \delta p_{j} 
      \delta p_{k} \right\rangle_{t} \label{Upsil3}
\end{eqnarray}

\noindent where the bracket $\langle\cdots\rangle_{t}$ with 
the subscript $t$ means the 
average over the probability density  $\rho(\delta\bfGamma,t)$ at time 
$t$: 
$\langle\cdots\rangle_{t}\equiv \int d\delta\bfGamma 
\rho(\delta\bfGamma,t)\cdots$.  
   These quantities consist of  
elements of the averaged matrix $\langle\delta\bfGamma 
\delta\bfGamma^{T} \rangle_{t}$, and satisfies the condition 
   
\begin{eqnarray}
   \Upsilon_{jk}^{(l)}(t)  = \Upsilon_{kj}^{(l)}(t).  
\label{SymmeUpsil}\end{eqnarray}

\noindent Equation (\ref{Fokke}) connects these quantities as 
   
\begin{eqnarray}
   \frac{d \Upsilon_{jk}^{(1)}(t)}{dt} &=& \frac{2}{m} 
       \Upsilon_{jk}^{(2)}(t)  \label{ConneUpsil1} \\
   \frac{d \Upsilon_{jk}^{(2)}(t)}{dt} &=& \frac{1}{m} 
      \Upsilon_{jk}^{(3)}(t) \label{ConneUpsil2} \\
   \frac{d \Upsilon_{jk}^{(3)}(t) }{dt} 
      &=& \sum_{\alpha=1}^{N}\sum_{\beta=1}^{N} 
   D_{j \alpha k \beta }Ÿ\Upsilon_{\alpha\beta}^{(1)}(t) 
\label{ConneUpsil3}\end{eqnarray}

\noindent where we assumed the probability density 
$\rho(\delta\bfGamma,t)$ to be zero at the boundary of the 
tangent space, and used Eqs. (\ref{DiffuConst1}), 
(\ref{DiffuConst2}) and (\ref{SymmeUpsil}) to 
derive Eq. (\ref{ConneUpsil3}). 

   Now we introduce a 4-th rank tensor 
$T_{j k ln}$ satisfying the condition  

\begin{eqnarray}
   \sum_{\alpha=1}^{N}\sum_{\beta=1}^{N} 
   T_{\alpha j k \beta} T_{\beta n l \alpha} = \delta_{j l}  
   \delta_{k n} 
\label{OrthoT1}\end{eqnarray}
	  
\noindent so that we obtain 

\begin{eqnarray}
   \sum_{\alpha=1}^{N}\sum_{\beta=1}^{N} 
   \sum_{\mu=1}^{N}\sum_{\nu=1}^{N}
   T_{j  \alpha \beta  k} D_{\alpha\mu\nu\beta} T_{n \nu \mu l} 
      = \Lambda_{j k} \delta_{j l} \delta_{k n}
\label{OrthoT2}\end{eqnarray}

\noindent with the real 2nd rank tensor $\Lambda_{jk}$ 
satisfying the condition 
$\Lambda_{jj} \geq 0$ \cite{note1}. 
   (Concerning Eqs. (\ref{OrthoT1}) and (\ref{OrthoT2}), for example, 
the existence of the real tensor $\Lambda_{jk}$ satisfying Eq. 
(\ref{OrthoT2}) simply comes from the fact that by the condition 
(\ref{DiffuConst2}) we can regard the quantity $D_{jkln}$ as the matrix 
element ${\cal D}_{\gamma_{1}(j,n)\gamma_{2}(k,l)}$ 
of the $N^{2} \times N^{2}$ real symmetric matrix 
${\cal D}\equiv({\cal D}_{\gamma_{1}(j,n)\gamma_{2}(k,l)})$ 
where $\gamma_{n}=\gamma_{n}(j,k)$, $n=1,2$ are 
functions from $j\in\{1,2,\cdots,N\}$ and 
$k\in\{1,2,\cdots,N\}$  to $\gamma_{n}\in\{1,2,\cdots,N^{2}\}$.)
   We will discuss an example of the tensors $T_{jkln}$ and $\Lambda_{jk}$ 
in the next section. 
   By using the tensor $T_{j k l n}$ we transform the quantities 
 $\Upsilon_{jk}^{(l)}(t)$ to $\tilde{\Upsilon}_{jk}^{(l)}(t)$ as 

\begin{eqnarray}
   \tilde{\Upsilon}_{jk}^{(l)}(t) 
   \equiv \sum_{\alpha=1}^{N}\sum_{\beta=1}^{N}
   T_{j\alpha\beta k} \Upsilon_{\alpha\beta}^{(l)}(t). 
\label{TransUpsil}\end{eqnarray}

\noindent  The inverse transformation to derive 
the quantity $\Upsilon_{jk}^{(l)}(t)$ from the quantity  
$\tilde{\Upsilon}_{\alpha\beta}^{(l)}(t)$ is simply given by 

\begin{eqnarray}
   \Upsilon_{jk}^{(l)}(t) 
   = \sum_{\alpha=1}^{N}\sum_{\beta=1}^{N}
   T_{\beta kj \alpha} \tilde{\Upsilon}_{\alpha\beta}^{(l)}(t)
\label{InverUpsil}\end{eqnarray} 

\noindent  noting the relation (\ref{OrthoT1}).
   Especially we should notice the relations  
$\langle |\delta \bfq |^{2} \rangle_{t} 
   = \sum_{\alpha=1}^{N}
   \tilde{\Upsilon}_{\alpha\alpha}^{(1)}(t)$ and 
$\langle |\delta \bfp |^{2} \rangle_{t} 
   = \sum_{\alpha=1}^{N}
   \tilde{\Upsilon}_{\alpha\alpha}^{(3)}(t)$  
under the condition $\sum_{\alpha=1}^{N} T_{j \alpha \alpha k} 
= \delta_{jk}$.

   We introduce the positive (or zero)   Lyapunov exponents 
$\lambda_{j}$ as 
   
\begin{eqnarray}
   \lambda_{j} = \lim_{t\rightarrow\infty} 
   \frac{1}{2t} \ln \frac{\tilde{\Upsilon}_{jj}^{(1)}(t)}
   {\tilde{\Upsilon}_{jj}^{(1)}(0)} , 
\label{Lyapu}\end{eqnarray} 

\noindent 
namely as the exponential rate of the time evolution 
of the average of magnitude of the 
infinitesimal deviation of the phase space orbit in the long time limit.
   As shown in Appendix \ref{LyapuRando},  these Lyapunov 
exponents are simply given by   

\begin{eqnarray}
   \lambda_{j} = \left[Ÿ\frac{\Lambda_{jj}}{(2m)^{2}} \right]^{1/3}Ÿ
\label{Lyapu2}\end{eqnarray} 

\noindent using the quantity $\Lambda_{jj}$ introduced in Eq. 
(\ref{OrthoT2}). 
   Equation (\ref{Lyapu2}) is the key result of this paper. 
   This equation connects the Lyapunov exponents directly with 
the tensor $D_{jkln}$ representing the strength of correlation 
of the particle interactions and potentials given by fixed scatterers, 
and also shows the fact that in the system described by the random matrix 
dynamics the Lyapunov exponents are independent of the initial 
condition like in the deterministic chaos. 
   It should be noted that the Lyapunov exponents 
$\lambda_{j} $ given by Eq. (\ref{Lyapu2}) are the same with the 
quantities derived from the equations $\lim_{t\rightarrow\infty} 
 (2t)^{-1}\ln [\tilde{\Upsilon}_{jj}^{(k)}(t)/\tilde{\Upsilon}_{jj}^{(k)}(0)]
, k=2,3$ in this approach (See Appendix \ref{LyapuRando}.).
 
   We sort the  Lyapunov exponents $\lambda_{1}$, $\lambda_{2}$, 
$\cdots$,  $\lambda_{N}$ so that they contain an decreasing (or equal) 
sequence,  and introduce the set 
$\{\lambda^{[1]}, \lambda^{[2]}, \cdots,
\lambda^{[N]} \}$ satisfying the condition 

\begin{eqnarray}
   \{\lambda^{[1]}, \lambda^{[2]}, \cdots, \lambda^{[N]}\} 
   = \{\lambda_1, \lambda_2, \cdots,\lambda_N\} 
\label{LyapuSpect1}\end{eqnarray}
\begin{eqnarray}
   \lambda^{[1]}\geq \lambda^{[2]} \geq \cdots \geq 
   \lambda^{[N]}.
\label{LyapuSpect2}\end{eqnarray}

\noindent This sorted set of the  Lyapunov exponents  is called the 
Lyapunov spectrum.


\section{A simplification}
\label{Sym} 

   Before considering Lyapunov spectra in concrete models 
using the formula given in the previous section, we discuss an assumption 
to simplify model calculations.   
   We consider the case where all matrix elements 
$R_{jk}(t)$, $j=1,2,\cdots,N$ and $k=1,2,\cdots,N$ have the same time 
dependence, namely 

\begin{eqnarray}
   R_{jk}(t) = r(t) A_{jk} 
\label{RandoAssum}\end{eqnarray}

\noindent where $r(t)$ is a normalized Gaussian white randomness in the 
sense of 
%
$   \langle r(t_{1})) r(t_{2})
$ $      \cdots r(t_{2n-1})\rangle = 0$ and  
$\langle r(t_{1})  r(t_{2}) 
$ $      \cdots r(t_{2n})
      \rangle $
$  = \sum_{P_{d}} 
$ $      \delta(t_{j_{1}}-t_{j_{2}})
      \delta(t_{j_{3}}-t_{j_{4}})
$ $      \cdots
$ $      \delta(t_{j_{2n-1}}-t_{j_{2n}})$
%
for any integer $n$ and $A\equiv(A_{jk})$ is a 
time-independent  $N \times N$ matrix. 
   The matrix $A$ must be a symmetric matrix:
   
\begin{eqnarray}
   A_{kj} = A_{jk}, 
\label{SimmeMatriA}\end{eqnarray}

\noindent because the matrix $R(t)$ is symmetric. 

   The assumption (\ref{RandoAssum}) simplifies our considerations, 
because in this case the tensor $D_{jkln}$ is represented 
as the multiplication of the matrix element $A_{jk}$ with  $A_{ln}$. Thus

\begin{eqnarray}
   D_{jkln}= A_{jk}A_{ln}
\label{TensorDbyA}\end{eqnarray}

\noindent and the conditions (\ref{DiffuConst1}) and 
(\ref{DiffuConst2}) for the tensor $D_{jkln}$  
are automatically satisfied under the condition (\ref{SimmeMatriA}).  
   The condition (\ref{SimmeMatriA}) also implies that the matrix $A$ 
is diagonalizable using an orthogonal matrix $U\equiv(U_{jk})$ 
satisfying $U^{T}U=UU^{T}=I$, namely
$ (U^{T} A U )_{jk} = a_{j}\delta_{jk}$ with real  
eigenvalues $a_{l}, l=1,2,\cdots,N$ of the matrix $A$. 
   The 4-th rank tensor $T_{jkln}$ is constructed as 
   
 \begin{eqnarray}
   T_{jkln} = U_{kj}ŸU_{ln} 
\label{TensorTbyU}\end{eqnarray}
 
 \noindent which satisfies the conditions (\ref{OrthoT1}), 
 (\ref{OrthoT2}) 
 and  $\sum_{\alpha=1}^{N}T_{j \alpha \alpha k} = \delta_{jk}$. 
    Here the 2-nd rank tensor $\Lambda_{jk}$ is represented as 
 
\begin{eqnarray}
    \Lambda_{jk} = a_{j} a_{k}
\label{Lambdbya}\end{eqnarray}

\noindent which satisfies the conditions $\Lambda_{jj}\geq 0$. 
   Equations (\ref{Lyapu2}) and (\ref{Lambdbya}) lead to the expression of 
the Lyapunov exponents as  
   
\begin{eqnarray}
   \lambda_{j} = \left|\frac{a_{j}}{2m}\right|^{2/3}Ÿ.
\label{Lyapu3}\end{eqnarray}
   
\noindent After all under the assumption (\ref{RandoAssum}) 
the calculation of the Lyapunov spectrum is 
attributed  to the eigenvalue problem of the matrix 
$A$.


\section{One Dimensional Models and 
Steps of their Lyapunov Spectra}

  In this section we consider simple one dimensional models 
and calculate their 
Lyapunov spectra using the formula given in Section \ref{Lya}  
under the assumptions discussed in Section \ref{Sym}. 
   We are generally interested in which ingredients 
of the model system lead to particular features of  
the Lyapunov spectra. 
   We are especially interested in models which satisfy total 
momentum conservation and show stepwise structures in
their Lyapunov spectra.
   Numerically the observation of Lyapunov modes is associated
with systems with a stepwise structure of their Lyapunov spectrum,
such as the many hard disk system of Refs. 
\cite{Mil98a,Mil98b,Del96,Pos00,Mcn01a}. 

   We construct a model consisting of $N$ particles in 
a one dimensional space.   
   In this case the off-diagonal matrix element $A_{jk}, j \neq k$ 
represents the strength of 
the interaction between the $j$-th particle and the $k$-th particle. 
   The diagonal matrix element $A_{jj}$ is determined by 
 the total momentum conservation which imposes the condition 

\begin{eqnarray}
   \sum_{k=1}^{N} A_{jk} = 0 
\label{MomenConse}\end{eqnarray}
 
\noindent for the matrix $A$. 
   Equation (\ref{MomenConse}) is derived from Eq. (\ref{RandoAssum}) and 
the equation $\sum_{k=1}^{N} R_{jk}(t)= 0$ satisfied at any time $t$. 
   Using the formula (\ref{Lyapu3}) this condition 
implies that the $N$-dimensional vector, whose components are equal, 
is an eigenstate of the matrix $A$ corresponding to the eigenvalue $0$, 
so we obtain

\begin{eqnarray}
   \lambda^{[N]} = 0,
\label{ZeroLyapu}\end{eqnarray}

\noindent namely there is a zero Lyapunov exponent corresponding 
to the conservation of total momentum.

\subsection{One dimensional model with a stepwise structure
of its Lyapunov spectrum}

   As a first step, we consider the case where each particle 
interacts only with its nearest neighbor particles with  
the same strength of interaction. 
   We impose periodic boundary conditions, namely  
that the particles are on a one dimensional ring structure. 
   This situation is described by the matrix $A
=A^{(0)} \equiv (A^{(0)}_{jk})$ defined by 
   
\begin{eqnarray}
   A^{(0)}_{jk} &\equiv&  \omega\left[
   -2\delta_{jk} +\delta_{j(k+1)} +\delta_{(j+1)k} \right. \nonumber \\
    && \hspace{1cm} \left.   +
   \delta_{j(k-N+1)} +\delta_{k(j-N+1)} \right]
\label{MatriA0}\end{eqnarray} 

\noindent with a (non-zero) real constant $\omega$. 
   We can calculate the eigenvalues of the matrix $A^{(0)}$ (See 
Appendix \ref{LyapuOneDimen} for the calculation details.), 
and obtain the Lyapunov exponents $\lambda_{n}=\lambda_{n}^{(0)}$ as  

\begin{eqnarray}
   \lambda_{n}^{(0)}= \left[\frac{|\omega|}{m}
   \left(1- \cos\frac{2\pi n}{N}\right)\right]^{2/3}.
\label{LyapuModel1}\end{eqnarray}

\noindent 
   It is important to note that the Lyapunov exponents given by Eq. 
(\ref{LyapuModel1}) satisfy the conditions $\lambda_{N}^{(0)} = 0$ and 

\begin{eqnarray}
   \lambda_{j}^{(0)}=\lambda_{N-j}^{(0)} 
\label{LyapuDegen}\end{eqnarray}

\noindent for $j<N/2$, namely the Lyapunov spectrum has 
degeneracies. In other words the spectrum has a stepwise structure. 
   The maximum Lyapunov exponent is given by $(2|\omega|/m)^{2/3}$ 
when $N$ is even  or 
$\{|\omega| [1+\cos(\pi/N)]/m\}^{2/3}$ when $N$ is odd.


\subsection{Robustness of the stepwise structure of 
the Lyapunov spectrum to a perturbation} 

Next we consider the case where each particle interacts 
with its nearest neighbor particles by slightly 
different interaction strengths.  
   This situation is described by the matrix $A=A^{(0)}
+\varepsilon A^{(1)}$ with $\varepsilon$ a small parameter 
and the matrix $A^{(1)}$ defined 
   
\begin{eqnarray}
   A^{(1)}_{jk} &=&
   -( \chi_{j} + \chi_{j+1} )\delta_{jk} +
   \chi_{j} \delta_{j(k+1)} +\chi_{k}\delta_{k(j+1)} 
   \nonumber \\
   && + 
    \chi_{1}Ÿ\left[\delta_{j(k-N+1)} 
   +\delta_{k(j-N+1)}  \right]
\label{MatriA1}\end{eqnarray}  
   
\noindent where $\chi_{j}$, $j=1,2,\cdots,N+1$ are real constants 
satisfying the condition $\chi_{N+1}=\chi_{1}$. 
   In this case we can calculate the eigenvalues of this matrix $A$ to first 
order in the small parameter $\varepsilon$, and obtain the 
Lyapunov exponents $\lambda_{n} = \lambda_{n}^{(0)} 
+ \varepsilon\lambda_{n}^{(1)} +{\cal O}(\varepsilon^{2})$ in the 
expanded form by the parameter $\varepsilon$ with the 
quantities 

\begin{eqnarray}
   \lambda_{j}^{(1)} &=& \frac{2\lambda_{j}^{(0)}}{3\omega} 
    \left(\tilde{\chi}_{0} + |\tilde{\chi}_{j}|\; \right) 
    \label{FirstOrderLyapu1}  \\
   \lambda_{N-j}^{(1)} &=& \frac{2\lambda_{N-j}^{(0)}}{3\omega}
   \left(\tilde{\chi}_{0} - |\tilde{\chi}_{j}|\; \right)
\label{FirstOrderLyapu2} \end{eqnarray} 

\noindent  in the case of  $j < N/2$, 
where $\tilde{\chi}_{j}$ is defined by 

\begin{eqnarray}
   \tilde{\chi}_{j}\equiv  \frac{1}{N} \sum_{\alpha=1}^{N} \chi_{\alpha} 
   \exp\left(\frac{4\pi i \alpha j}{N}\right). 
 \label{TildeChi} \end{eqnarray} 

\noindent (See Appendix \ref{LyapuOneDimen} for the derivations of 
Eqs. (\ref{FirstOrderLyapu1}) and (\ref{FirstOrderLyapu2}).)
   Here we numbered so that we obtain  
$\omega(\lambda_{j}^{(1)} - \lambda_{N-j}^{(1)}) \geq 0$.
   It may be noted  that the quantity 
$\tilde{\chi}_{0}$ in Eqs. (\ref{FirstOrderLyapu1}) and 
(\ref{FirstOrderLyapu2}) is simply the arithmetic average of the coefficients 
$\chi_{j}$, $j=1,2,\cdots,N$: $\tilde{\chi}_{0}=N^{-1}\sum_{j=1}^{N} 
\chi_{j}$.
   We also obtain  $\lambda_{N/2}^{(1)} = 2  
 \lambda_{N/2}^{(0)}\tilde{\chi}_{0}/(3\omega)$
in the case of the number $N$ to be even, and  $\lambda_{N}^{(1)}=0$ 
for the 
first order correction to the non-degenerate Lyapunov exponents.  

   Equations (\ref{FirstOrderLyapu1}) and (\ref{FirstOrderLyapu2}) tell us 
about robustness of the stepwise structure of the Lyapunov spectrum 
appearing in the non-perturbed system against the perturbation. 
   To discuss this point we consider the deviation 
$ \left| \lambda_{j} - \lambda_{N-j} \right|$
 of the two points in  the $j$-th step of the 
Lyapunov spectrum, which is given by 

\begin{eqnarray}
   \left| \lambda_{j} - \lambda_{N-j} \right|
   =\frac{4}{3}\left|\frac{\varepsilon \tilde{\chi}_{j}}{\omega}\right|
    \lambda_{j} \; 
    + {\cal O}(\varepsilon^{2})
\label{DeviaStep} \end{eqnarray}

\noindent  in the case of $j < N/2$. 
   Equation (\ref{DeviaStep}) implies that the degeneracies, namely the 
stepwise structure of the Lyapunov spectrum is removed by the 
perturbation $\varepsilon A^{(1)}$, but their deviation are small 
in a region of small Lyapunov exponents  rather than in a region 
of large Lyapunov exponents, as far as the quantity 
$\tilde{\chi}_{j}$ is almost $j$-independent. 
   In other words, this consideration suggests that the stepwise 
structure of the Lyapunov spectrum is robust in the region of small Lyapunov 
exponents. 
   It should be emphasized that this is consistent with numerical 
results of the Lyapunov spectrum in a many hard disk model, 
which shows the stepwise structure of the Lyapunov spectrum 
only in a region of small Lyapunov exponents 
\cite{Mil98a,Mil98b,Del96,Pos00,Mcn01a}.  
    By using Eqs. (\ref{FirstOrderLyapu1}) and 
(\ref{FirstOrderLyapu2})  we can also  discuss a perturbational effect 
in a global shape of the Lyapunov spectrum.
   Eqs. (\ref{FirstOrderLyapu1}) and (\ref{FirstOrderLyapu2}) 
lead to the shift of the  $j$-th step of the Lyapunov spectrum as  

\begin{eqnarray}
   \frac{\lambda_{j} + \lambda_{N-j}}{2} 
   - \lambda_{j}^{(0)} =  \frac{2\tilde{\chi}_{0}}{3\omega}
    \lambda_{j} \; 
    \varepsilon + {\cal O}(\varepsilon^{2})
\label{SiftStep} \end{eqnarray}

\noindent  in the case $j < N/2$. 
   Equation (\ref{SiftStep}) implies that 
in the case of  $\tilde{\chi}_{0}\varepsilon/\omega>0$ 
($\tilde{\chi}_{0}\varepsilon/\omega<0$) 
the perturbation 
$\varepsilon A^{(1)}$ makes the slope of the Lyapunov spectrum more  
(less) steep than in the non-perturbed case.

\vspace{0.6cm}
\begin{figure}[t]
   \epsfxsize 7.0cm 
   \centerline{\epsffile{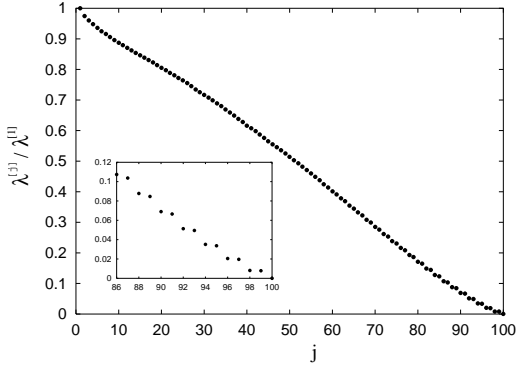}}
   \vspace{0.4cm}
   \caption{\lines Lyapunov spectra of a one dimensional system with nearest 
   neighbor interactions in the case of randomly chosen matrix elements. 
   The inset: Enlarged graph of a small Lyapunov exponent region.} 
\label{1Dnear} 
\end{figure}  
\vspace{0.2cm}   

   Now we discuss the robustness of the stepwise structure of the 
Lyapunov spectrum against perturbation 
in a little different way, in which the fluctuations of the 
matrix elements of $A$ are large enough so that the above 
first order perturbed discussion is no longer correct. 
    We consider a one dimensional system described 
by the matrix $A = A^{(1)}$ given by Eq. (\ref{MatriA1}), 
in which the quantities $\chi_{j}$, $j=1,2,\cdots,N$ 
are chosen randomly from the region $(a,b)$. 
    Figure \ref{1Dnear} is a Lyapunov 
spectrum normalized by the maximum Lyapunov exponent in such a system 
consisting of 100 particles ($N=100$).  
    Here, we chose the region $(a,b)$ as $a=\omega/2$ and $b=3\omega/2$, 
and took the arithmetic average over this randomness for the normalized 
Lyapunov spectrum. 
   Although the magnitude $b-a \;(=\omega)$ of fluctuations in the matrix 
elements of the matrix $A$ is the same scale as the averaged magnitude 
$(a+b)/2 \;(=\omega)$ of the matrix elements of the matrix $A$, we can 
still recognize some steps in 
the Lyapunov spectrum in the region of small Lyapunov exponents.     
   It is interesting to note that in such a case the global 
shape of the Lyapunov spectrum is rather close to a straight 
line, like discussed in Refs. \cite{Liv86,Liv87,Pal86,New86,Eck88}.


\subsection{Effect of long range interactions and wave like 
structures in eigenstates}
 
   In this subsection we consider the effects of long range interactions 
between particles on the Lyapunov spectrum. 
   The effect of the long range interactions between particles can be taken 
into account using the matrix $A=A^{[n_{l}]}\equiv (A_{jk}^{[n_{l}]})$ 
defined by  

\begin{eqnarray}
    A_{jk}^{[n_{l}]} &=& \sum_{l=1}^{n_{l}} \left[
   -( \sigma_{j}^{[l]} + \sigma_{j+l}^{[l]})\delta_{jk} +
    \sigma_{j}^{[l]} \delta_{j(k+l)} 
      +\sigma_{k}^{[l]}\delta_{k(j+l)} 
      \right. \nonumber \\
   && + \left.
    \sigma_{j-N+l}^{[l]} \delta_{j(k+N-l)}
   +    \sigma_{k-N+l}^{[l]} \delta_{k(j+N-l)}  \right] 
\label{MatriA2}\end{eqnarray} 

\noindent with real constants $\sigma_{j}^{[l]}$, 
where $n_{l} (<N/2)$ is the length of the interactions and its 
value means that particles can interact with up to their $n_{l}$-th 
nearest neighbor particles. 
   As shown in Appendix \ref{LyapuOneDimenLong}, we can obtain 
an analytical expression for the Lyapunov spectrum derived from 
the matrix (\ref{MatriA2}) if all of the non-zero matrix elements 
$\sigma_{j}^{[l]}$, $j=1,2,\cdots,N$, $l=1,2,\cdots,n_{l}$ 
take the same value.

\vspace{0.6cm}
\begin{figure}[t]
   \epsfxsize 7.0cm 
   \centerline{\epsffile{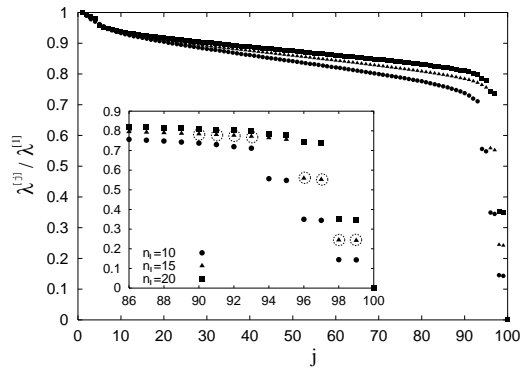}}
   \vspace{0.4cm}
   \vspace{0cm}
   \caption{\lines Lyapunov spectra of one dimensional systems with 
      different length $n_{l}$ of long range interactions 
	  in the case of $N=100$. 
         The three graphs correspond 
      to the cases $n_{l}=10$ (the circle dots), $n_{l}=15$ 
      (the triangle dots) and  $n_{l}=20$ (the square dots), 
      respectively. 
   The inset: Enlarged graphs of a small Lyapunov exponent region.}
\label{1Dlong} 
\end{figure}  
\vspace{0.2cm}

   In this subsection we consider the case where the matrix element 
$\sigma_{j}^{[l]}$ is chosen randomly, like in the model 
discussed in the end of the previous subsection.   
   Figure \ref{1Dlong} is the Lyapunov spectrum  normalized 
by the maximum Lyapunov exponent in the system described by the 
matrix  $A=A^{[n_{l}]}$ with the quantities $\sigma_{j}^{[l]}$ 
chosen randomly from the region $(\omega/2,3\omega/2)$ for 
the case of $n_{l}=10$, $15$ and $20$ in the system consisting of 100 
particles (N=100).  
   (This randomness in the quantities $\sigma_{j}^{[l]}$ 
is adopted to draw all of the figures hereafter in this subsection.)
   In these graphs we took the arithmetic average of the 
Lyapunov spectra over 
the randomness of the quantities $\sigma_{j}^{[l]}$. 
   This figure shows that the long range interactions 
separates the Lyapunov spectrum clearly 
into a part exhibiting stepwise structure and a part changing smoothly. 
   (The randomness of the quantities $\sigma_{j}^{[l]}$ is not 
essential for this separation in the Lyapunov spectrum. 
   It mainly plays the role of smoothing the Lyapunov spectrum 
in the region of large Lyapunov exponents.)
   The stepwise structure of the Lyapunov spectrum appears only 
in a region of small Lyapunov exponents, as suggested in the 
previous subsection.
 
\vspace{0.6cm}
\begin{figure}[t]
   \epsfxsize 7.0cm 
   \centerline{\epsffile{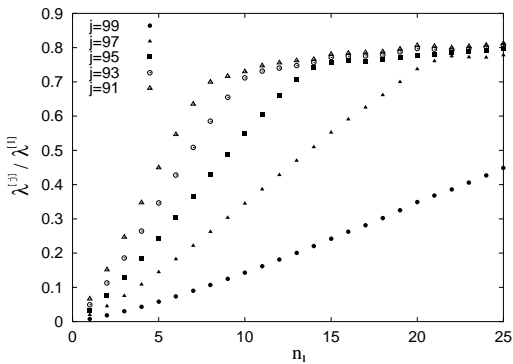}}
   \vspace{0.4cm}
   \caption{\lines Dependence of the long range interaction length $n_{l}$ 
   on the Lyapunov exponents $\lambda^{[j]}$ in the case of $N=100$.
   The graphs with the circle dots, the triangle dots, the square 
   dots, the open circle dots and the open triangle dots  
   correspond to the case of $j=99,97,95,93$ and $91$, 
   respectively.}
\label{1Dstep1} 
\end{figure}  
\vspace{0.2cm}

   The stepwise region of the Lyapunov spectrum 
is smaller and smaller as  the interaction length 
$n_{l}$ of particles is longer and longer.  
 (As a limit of the long range interactions, as shown 
in Appendix \ref{LyapuOneDimenLong}, we can easily show 
that in the system, where each particle interacts 
with all the other particles with the same strength, 
all of the positive Lyapunov exponents take the same value.)
   Besides, the heights of the steps of the Lyapunov spectrum 
are approximately proportional to the length $n_{l}$ of the interactions. 
   These characteristics are clearly evident in Fig. \ref{1Dstep1}, 
showing the length $n_{l}$ dependence of the $j$-th 
positive Lyapunov exponents $\lambda^{[j]}$, $j=91,93,95,97,99$ 
in the case of 100 particles. 
   In these graphs we also took the arithmetic average 
of the Lyapunov exponents over the randomness 
of the quantities $\sigma_{j}^{[l]}$. 
   This figure shows that the values of small Lyapunov exponents 
increase with the interaction length $n_{l}$ as far as they are in 
the region of the steps of the Lyapunov spectrum, and 
if the interaction length $n_{l}$ is bigger than a critical value then 
their $n_{l}$ 
dependences are slowed down, meaning that they are in the 
region of the Lyapunov spectrum that is changing smoothly.

\begin{figure}[t]
   \epsfxsize 8.6cm 
   \centerline{\epsffile{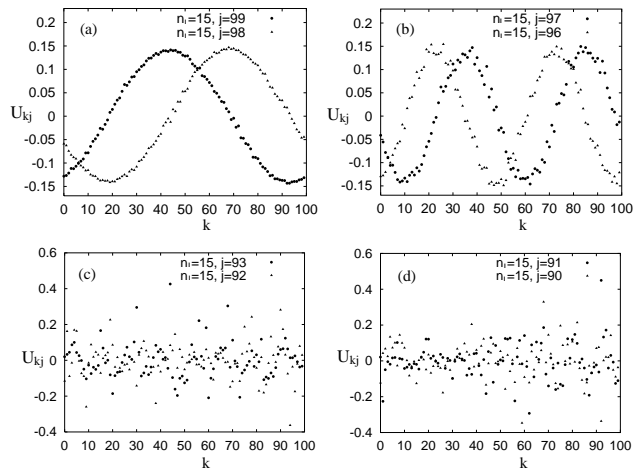}}
   \vspace{0.4cm}
   \caption{\lines Eigenstates of matrix $A^{[n_{l}]}$ in the case 
      of $n_{l}=15$ and $N=100$:  
      the graphs (a) ($j=98,99$), the graph (b) ($j=96,97$), 
      the graph (c) ($j=92,93$) and the graph (d) ($j=90,91$)  
      for the Lyapunov exponent $\lambda^{[j]}$.
      These Lyapunov exponents correspond to the triangle 
      dots surrounded by the broken lines in the inset of 
      Fig. \ref{1Dlong}.}
\label{1Deigen1} 
\end{figure}

   It may be interesting to investigate the difference between
the two parts of the Lyapunov spectrum from the 
point of view of the eigenstate. 
   Figure \ref{1Deigen1} is the real eigenstate of the matrix 
$A^{[n_{l}]}$ in the case of $n_{l}=15$ and $N=100$ 
with randomly chosen quantities $\sigma_{j}^{[l]}$.  
   (Note that we did not take the average 
over the randomness of the quantities $\sigma_{j}^{[l]}$ 
to draw this figure.) 
   The graphs (a), (b), (c) and (d) are the eigenstates 
corresponding to the Lyapunov exponents $\lambda^{[j]}$ for $j=99, 98$, 
$j=97, 96$, $j=93, 92$ and $j=91,90$, respectively. 
   These Lyapunov exponents correspond to the 
triangle dots surrounded by the broken lines in the inset  
of Fig. \ref{1Dlong}. 
   We can see clear wave like structures of approximately sinusoidal type 
in the graphs (a) and (b), which correspond to the 
Lyapunov exponents composing the steps in the Lyapunov spectrum. 
   The wavelength of the waves in the graph (b) is half of the 
wavelength of the waves in the graph (a). 
   On the other hand, we cannot recognize such a 
wave like structure in the graphs (c) and (d), 
which belongs to the Lyapunov exponents in the part of 
the Lyapunov spectrum which is changing smoothly. 

\vspace{0.6cm}
\vspace{0.2cm}

   It should be noted that the wave like structure of the eigenstates 
already appears even in the system described by the matrix 
$A=A^{(0)}$ defined by Eq. (\ref{MatriA0}) with nearest 
neighbor interactions of a constant strength. 
   Therefore the point is that such a wave like structure of the eigenstates 
is not destroyed by the long range interactions and the random interaction 
strengths only in the eigenstates which 
correspond to the small positive Lyapunov exponents in the 
steps of the Lyapunov spectrum.
  
   One may regard the long range interactions discussed 
in this subsection as a kind of high dimensional effect, 
because if particles move in a two or three dimensional space 
then they can interact with more than two particles, 
even in hard-core interactions.
   However, some high dimensional effects, for example, the total 
momentum conservation in each orthogonal direction and so on, are 
still missing in the discussions in this subsection.   
   In the next section we consider high dimensional   
effects more explicitly.




\section{Two Dimensional Model with a Triangle Lattice Structure}

   In the one dimensional models discussed in the previous section, 
the steps of the Lyapunov spectrum are caused by the periodic
boundary conditions, and all of them consists of 2 only points. 
   In this section we consider a two dimensional model, and show  
that it is possible to get wider steps than in the one dimensional 
model as a high dimensional effect. 
   Specifically, we concentrate on a model which has 
steps consisting of 4 and 8 points, which are 
actually found in a square system consisting of many hard 
disks \cite{Pos00}. 

   In the two dimensional system, the position of each particle
is specified by a two dimensional vector $\bfq_{j}$, $j=1,2,\cdots,
\tilde{N}$, where $\tilde{N}$ is the number of the particles and 
$N=2\tilde{N}$. 
   In such a case the matrix $A$ is a $(2\tilde{N}) \times 
(2\tilde{N})$ matrix, and can be represented as a block matrix 
consisting of $2 \times 2$ matrices $B^{(jk)}$, 
$j=1,2,\cdots,\tilde{N}$ and $k=1,2,\cdots,\tilde{N}$.  
   Here $B^{(jk)}\equiv(B_{j'k'}^{(jk)})$ is given by 

\begin{eqnarray}
   B^{(jk)} &\equiv& \left(
   \begin{array}{cc}
      A_{(2j-1)(2k-1)} & A_{(2j-1)(2k)} \\
      A_{(2j)(2k-1)} & A_{(2j)(2k)} 
   \end{array}
   \right) 
\label{2DBlock}\end{eqnarray}

\noindent corresponding to the matrix $-\partial^{2} V  
/ \partial \bfq_{j}\partial \bfq_{k}$, and its components  
represent the strengths of the interactions between 
components of positions of the $j$-th particle and the 
$k$-th particle.
   The matrix $B^{(jk)}$ is symmetric and the matrix $A$ is also  
symmetric in the sense of the block matrix, namely 
 
\begin{eqnarray}
   B_{j'k'}^{(jk)} = B_{k'j'}^{(jk)} = B_{j'k'}^{(kj)}. 
\label{2DBlockSymme}\end{eqnarray}
 
\noindent The diagonal block $B^{(jj)}$ is determined by the condition
of the total momentum conservation 

 \begin{eqnarray}
   \sum_{k=1}^{\tilde{N}} B^{(jk)} = 0, 
\label{2DMumenConser}\end{eqnarray}

\noindent which comes from  Eq. (\ref{RandoAssum}) and the relation 
$\sum_{k=1}^{\tilde{N}} \partial^{2} V 
/ \partial \bfq_{j}\partial \bfq_{k} = 0$ 
and is simply the two dimensional version of the condition 
(\ref{MomenConse}). 
   This condition (\ref{2DMumenConser}) implies that 
the Lyapunov exponents include at least two zero components:  

\begin{eqnarray}
    \lambda^{[N-1]} = \lambda^{[N]} = 0  
\label{ZeroLyapu2}\end{eqnarray}

\noindent corresponding to the conservation of the total momentum.

\begin{figure}[t]
   \epsfxsize 8.6cm 
   \centerline{\epsffile{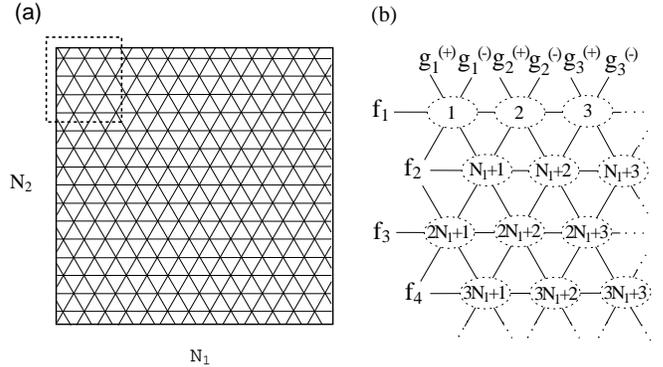}}
   \vspace{0.4cm}
   \caption{\lines 
   (a) The triangular lattice system with a square shape. 
   $N_{1}$ is the number of the particles in a horizontal 
   line, and $N_{2}$ is the number of the horizontal lines.   
   (b) Numbering of the particles and the boundary condition
   in the part surrounded by the square of broken line 
   in the figure (a).}
\label{2Dtria} 
\end{figure}

   To construct a two dimensional model it is convenient to use 
a lattice picture, because in the approach of this paper, under 
the assumption discussed in Section \ref{Sym}, the model 
is specified by pairs of interacting particles and 
their interaction strengths. 
   In such a lattice, each lattice site corresponds to 
a particle and a connection between sites means that particles 
on those two sites  
can interact with each other as nearest neighbor particles.   
   Now we consider an equilateral triangular lattice 
from such a point of view. 
   This situation is motivated by the fact that  
in Ref. \cite{Pos00} the system consists of many disks which pack in a 
two dimensional space as a hexagonal close-packed structure in 
the high density limit. 
   The triangular lattice has three directions to connect the sites, 
and arrange the triangle lattice so that it fits within a square where one of 
sides of the square is parallel to one of the three directions of 
the triangle lattice, as shown in Fig \ref{2Dtria}(a) where 
such parallel lines are horizontal. 
   For simplicity we assume that the number of lattice sites 
in each horizontal line is equal and is given by $N_{1} (>1)$. 
   We put the number of such lines as $N_{2}$, which should be roughly 
given by $N_{2}\approx (2/\sqrt{3})N_{1}$. 

\vspace{0.6cm}
\vspace{0.2cm}

   Next, we introduce the boundary condition for this triangle 
lattice model. 
   For this purpose we assign numbers $1,2,\cdots,
N_{1}$ to the lattice sites (namely particles) 
in the first horizontal row of particles 
from left to right, and numbers $N_{1}+1,N_{1}+2,\cdots,2N_{1}$
in the second row and so on until the last row numbered
$(N_{2}-1)N_{1}+1,(N_{2}-1)N_{1}+2,\cdots, N_{2}N_{1}$, 
as shown in Fig. \ref{2Dtria}(b).   
   We define the number $f_{2j-1}$ as the 
number of the particle which do nearest neighbor interactions 
with the $[2(j-1)N_{1}+1]$-th particle, and the number $f_{2j}$ 
as the number of the particle which do nearest-neighbor 
interacts with the $[2(j-1)N_{1}+1]$-th, $[(2j-1)N_{1}+1]$-th and 
$[2jN_{1}+1]$-th particles ($j=1,2,\cdots,\mbox{Int}\{N_{2}/2\}$). 
   Here, $\mbox{Int}\{x\}$ means to take the integer 
part of $x$ for any real number $x$.  
   We also define the numbers $g_{j}^{(+)}$ and $g_{j}^{(-)}$ 
as the number of particles which do nearest neighbor interactions 
with the $j$-th particle through the upper-left line and the upper-right 
line, respectively ($j=1,2,\cdots,N_{1}$).
   (See Fig. \ref{2Dtria}(b) for the definitions of the numbers 
$f_{j}$ and $g_{j}^{(\pm)}$.)
   The number $f_{j}$ specifies the interaction between the 
left side and the right side of the square, and the numbers 
$g_{j}^{(\pm)}$ specifies the interaction between the upper 
side line and the lower side of the square. 
   In this section we consider the case that  
these numbers are given by 

\begin{eqnarray}
   f_{j} = j N_{1}
\label{Perio1}\end{eqnarray}
\begin{eqnarray}
   g_{j}^{(\pm)} = (N_{1}-1)N_{2} +  
   \min_{k\in\{1,2,\cdots\}} \left\{ h_{jk}^{(\pm)}; \; 
   h_{jk}^{(\pm)} > 0\right\}
\label{Perio2}\end{eqnarray}

\noindent where $h_{jk}^{(\pm)}$ is defined by 

\begin{eqnarray}
   h_{jk}^{(\pm)} = j \pm \mbox{Int}\left\{\frac{1}{2}
   \left(N_{2} - \frac{1\pm 1}{2}\right)\right\} 
   \mp kN_{1}.  
\label{FunctH}\end{eqnarray}

\noindent The boundary conditions (\ref{Perio1}) 
and (\ref{Perio2}) imply that every line of the three directions 
in the triangle lattice is periodic.    
   We assume the condition $g_{j}^{(+)}\neq g_{j}^{(-)}$ 
so that every particle has the six nearest-neighbor 
particles. 

   Before giving the matrix $A$ of the 
triangle lattice model including a long range interaction, 
we consider the matrix $A=\tilde{A}_{0}$ including 
only the effect of the nearest neighbor interactions 
of a constant strength.   
   We define the $2N_{1} \times 2N_{1}$ matrices $C_{l}$, 
$l=1,2,3,4$, each of which is a block matrix consisting 
of $2 \times 2$ matrices $C_{l}^{(jk)}$, $j=1,2,\cdots,N_{1}$, 
$k=1,2,\cdots,N_{1}$, by 

\begin{eqnarray}
   C_{l}^{(jk)} &\equiv&
   - 6{\cal B} \delta_{jk} +
   {\cal B} \left[ \delta_{j(k+1)} + \delta_{k(j+1)}\right.  
   \nonumber \\
   && \hspace{2cm}+ 
   \left. \delta_{j(k-N_{1}+1)} 
   +\delta_{k(j-N_{1}+1)}  \right] 
\label{MatriC1}\end{eqnarray} 
\begin{eqnarray}
   C_{2}^{(jk)} \equiv
   {\cal B} \left [\delta_{jk} 
   + \delta_{j(k+1)} + \delta_{j(k-N_{1}+1)} \right]  
\label{MatriC2}\end{eqnarray} 
\begin{eqnarray}
   C_{3}^{(jk)} \equiv 
   {\cal B} \left[ \delta_{jk} 
   + \delta_{k(j+1)} + \delta_{k(j-N_{1}+1)} \right]  
\label{MatriC3}\end{eqnarray}
\begin{eqnarray}
   C_{4}^{(jk)} \equiv 
   {\cal B} \left\{ \delta_{k\,[g_{j}^{(+)} - (N_{1}-1)N_{2}]} 
   + \delta_{k\,[g_{j}^{(-)} - (N_{1}-1)N_{2}]} \right\}  
\label{MatriC4}\end{eqnarray}

\noindent where ${\cal B}$ is a $2\times 2$ matrix.
   The matrix $\tilde{A}_{0}$ is introduced as the block matrix 
defined by 

\begin{eqnarray}
   \tilde{A}_{0} =  
   \left(\begin{array}{cccccccc}
      C_{1} & C_{2} & \overline{0} & \overline{0} & \overline{0}
             & \cdots & \overline{0} & C_{4} \\
      C_{2}^{T} & C_{1} & C_{3} & \overline{0} & \overline{0} 
             & \cdots & \overline{0} & \overline{0}     \\
      \overline{0} & C_{3}^{T} & C_{1} & C_{2} & \overline{0}
             &   &   &       \\
      \overline{0} & \overline{0} & C_{2}^{T}& C_{1} & C_{3} 
             &   &   &       \\
      \overline{0} & \overline{0} & \overline{0} & C_{3}^{T} & C_{1}    
             & \ddots &    &       \\
      \vdots & \vdots  &    &    &  \ddots 
             & \ddots &    &         \\
      \overline{0} & \overline{0} &    &    &      
             &   & C_{1} & C_{\tilde{\nu}} \\
      C_{4}^{T} & \overline{0} &       &    &       
             &   & C_{\tilde{\nu}}^{T} & C_{1}  
   \end{array}\right)
\label{MatriTA0}\end{eqnarray}

\noindent with the $2N_{1}\times 2N_{1}$ null  
matrix $\overline{0}$, where $\tilde{\nu}$ is $2$ ($3$) if $N_{2}$ is 
an odd (even) number.   

   The matrix $A=\tilde{A}^{[n_{l}]}$ of the triangle lattice model 
including the effect of the long range interactions up to 
the $n_{l}$-th nearest neighbor interactions is simply given as follows. 
   We consider the matrix $\tilde{{\cal A}}_{0}'$ 
given by the matrix $\tilde{A}_{0}$ except  
that the matrix ${\cal B}$ in the matrix 
$\tilde{A}_{0}$ are replaced by the $2\times 2$ matrices whose 
matrix elements are positive constants. 
   By using such a $(2N_{1}N_{2}) \times (2N_{1}N_{2})$ matrix 
$\tilde{{\cal A}}_{0}'$ we calculate the $n_{l}$ times 
multiplication $(\tilde{{\cal A}}_{0}')^{n_{l}}$ of 
the matrix $\tilde{{\cal A}}_{0}'$. 
   The non-zero elements of the matrix $\tilde{A}^{[n_{l}]}$ are 
equivalent to the non-zero elements of the matrix 
$(\tilde{{\cal A}}_{0}')^{n_{l}}$. 
   After determining the non-null 
upper off-diagonal blocks $B^{(jk)}$, $j<k$ 
of the matrix $\tilde{A}^{[n_{l}]}$ satisfying 
the condition (\ref{2DBlockSymme}) by 
such a process, the diagonal block $B^{(jj)}$ and 
the lower off-diagonal bocks $B^{(jk)}$, $j>k$ of 
the matrix $\tilde{A}^{[n_{l}]}$ 
is determined so that the matrix $\tilde{A}^{[n_{l}]}$ is symmetric 
and satisfies the condition (\ref{2DMumenConser}). 
   It should be noted that we could use a similar method to obtain   
the matrix $A^{[n_{l}]}$ including the effect of long range interactions 
in the one dimensional model in the previous section.

   We restrict our consideration in the case that 
the $2\times 2$ matrix $B^{(jk)}$, which 
is the block element of the matrix $A$ in the two dimensional 
system, is diagonalized.

\begin{eqnarray}
   B^{(jk)}_{12}=B^{(jk)}_{21}=0 
\label{Offdi}\end{eqnarray}

\noindent This means that two components of the position of each particle
do not interact with each other. 

\begin{figure}[t]
   \epsfxsize 7.0cm 
   \centerline{\epsffile{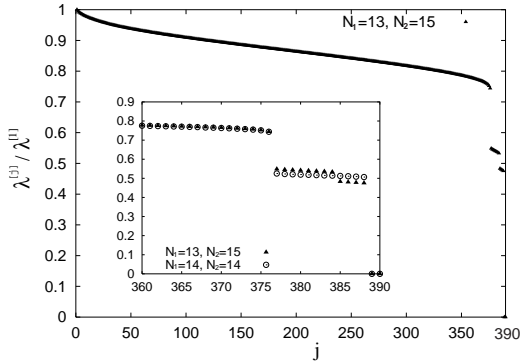}}
   \vspace{0.4cm}
   \caption{\lines Lyapunov spectrum of the triangle lattice system in the case 
   of $N_{1}=13$, $N_{2}=15\approx (2/\sqrt{3})N_{1}$ and 
   $n_{l}=3$ (the triangle dots).
   The inset:  Enlarged graphs of the part including the 
   stepwise structures in the Lyapunov spectra in the case of 
   $N_{1}=13$ and $N_{2}=15$  (the triangle dots) and the
   case of $N_{1}=N_{2}=14$ (the circle dots) for a comparison. 
   The Lyapunov spectrum in the case of $N_{1}=N_{2}=14$ is shifted 
   by $-2$ in the $j$-direction, so that the positions of the 
   zero Lyapunov exponents coincide in both the cases.}
\label{2Dlong} 
\end{figure}

   Now we calculate the Lyapunov spectrum for such a matrix 
$\tilde{A}^{[n_{l}]}$ by using the formula (\ref{Lyapu3}).
   Figure \ref{2Dlong} is the Lyapunov spectrum normalized 
by the maximum Lyapunov exponent in such a triangle 
lattice system in the case of 
$N_{1}=13$, $N_{2}=15 \approx (2/\sqrt{3})N_{1}=15.01110\cdots$ and
$n_{l}=3$. 
   Here we chose the non-zero elements in the 
upper triangle of the matrix $\tilde{A}^{[n_{l}]}$ randomly from 
the positive region $(0.2\tilde{\omega},1.8\tilde{\omega})$ with 
a (non-zero) real constant
$\tilde{\omega}$, and took the arithmetic average over this randomness. 
   As in the one dimensional model including the long-range 
interactions, the Lyapunov spectrum is separated into the part 
changing smoothly and the part having a stepwise structure, 
which appears in a region of small positive Lyapunov exponents. 

   A remarkable point in the two dimensional triangle lattice model, 
which the one dimensional models do not have, is the wide steps of 
the Lyapunov spectrum, especially the step consisting of 4 points 
and the step consisting of 8 points shown in Fig. \ref{2Dlong}. 
   To give an explanation for these wide steps, we should notice 
that in this model we set-up a periodic boundary condition for 
each of the three directions of the triangle lattice. 
   This boundary condition for each of the three directions 
can cause degeneracies in the Lyapunov 
spectrum like in the one dimensional cases. 
   Besides, in this model each site of the triangle lattice 
have two independent degrees of freedom corresponding to 
the two dimensionality of the particles, which can also cause 
degeneracy in the Lyapunov spectrum 
due to the condition (\ref{Offdi}).  
   Therefore we can get the step consisting of 4 points 
corresponding to each direction of the triangle lattice. 
   Moreover two of the three directions (namely the two directions 
other than the horizontal direction in Fig. \ref{2Dtria})  
have the same numbers of the sites in themselves, so they  
can cause a degeneracy in the Lyapunov spectrum.    
   After all we get the step consisting of 4 points 
corresponding to the horizontal direction, and the step 
consisting of 8 points corresponding to the other two directions.  
   This explanation is partly justified by the fact that 
the Lyapunov spectrum in the case of $N_{1}=N_{2}$ has the step 
consisting of 12 points in the Lyapunov spectrum, 
shown in the inset in Fig. \ref{2Dlong}, 
where we gave the averaged Lyapunov spectrum 
normalized by the maximum Lyapunov exponent in the 
case of $N_{1}=N_{2}=14$ and $n_{l}=3$.  
   Here, except for the numbers $N_{1}$ and $N_{2}$ we 
used the same boundary condition and the same randomness for 
non-zero elements of the matrix $\tilde{A}^{[n_{l}]}$ with in the 
case of $N_{1}=13$ and $N_{2}=15$ in Fig. \ref{2Dlong}. 
   This Lyapunov spectrum is shifted by $-2$ in the 
$j$-direction, so that the positions of the zero Lyapunov 
exponents coincide with the cases of $N_{1}=13$ and $N_{2}=15$. 
   One may also notice that the parts changing smoothly in 
the Lyapunov spectrum are almost indistinguishable in both 
the cases in the inset of Fig. \ref{2Dlong}. 

\vspace{0.6cm}
\vspace{0.2cm}

   Like in the one dimensional model, longer range interactions 
lead to a shorter region of stepwise structure in the Lyapunov spectrum 
in the two dimensional triangle lattice model. 
   On the other hand, a numerical simulation of the many disk system 
showed that 
the region of the stepwise structure depends on the 
aspect ratio of the rectangular system \cite{Pos00}. 
   This fact suggests that the long range interactions   
in the random matrix approach should depend on 
the aspect ratio of the rectangular system. 

   It should be emphasized that such wide step consisting of 
4 points or 8 points are actually found in the two dimensional 
deterministic chaotic system consisting of many hard-core disks 
numerically \cite{Pos00}. 
   Beside, we should also notice that in Fig. \ref{2Dlong} 
(and Fig. \ref{1Dlong}) a warp of the Lyapunov spectrum 
appears in a region of large Lyapunov exponents, which is 
also a characteristic of the system consisting of many 
hard-core disks.    
   However there are some points which the triangle lattice 
model in this section cannot give enough explanation when compared
with the numerically observed features of the Lyapunov spectrum for the 
deterministic two dimensional hard disk system. 
   First, in the numerical simulation of the hard-disk system 
each particle can interact 
with almost any other particle in the long time limit. 
   This should correspond to a big number of 
$n_{l} \approx N_{1}N_{2}$ in the triangle lattice model, 
but if we adopt such large number for $n_{l}$ then the stepwise 
structure disappears in the triangle lattice model. 
   In this sense, the model in this section may correspond to 
the high density case, in which each particle 
mainly interacts with only a few particles surrounding it.     
   Second, in the hard-disk system with a square shape 
the stepwise structure of the Lyapunov spectrum seems to be  
a repetition of the step consisting of 4 points and 
the step consisting of 8 points.
   On the other hand, in the triangle lattice model 
such a repetition of steps cannot be guaranteed. 
   As the third point, in the triangle lattice model 
in this section we adopt the boundary condition 
to make each of the three directions of the triangle 
lattice periodically, rather than the periodic boundary 
condition to make the up-side (the left-side) and the 
down-side (the right-side) of the square equivalently, 
which was adopted in the numerical simulation of 
the hard-disk system.


\section{Conclusion and Remarks}
 
  In this paper we have discussed the Lyapunov spectrum in many particle 
systems described by a random matrix dynamics. 
   We started from the many particle Hamiltonian mechanics without a
magnetic field, and introduced Gaussian white random interactions between 
the particles. 
   In such a system the dynamics of the tangent space is expressed by a 
Fokker-Planck equation, which leads to a direct connection between 
the positive (and zero) Lyapunov exponents and the time correlation 
of the matrix specifying the particles interactions.
   Using this formula, we calculated concretely the Lyapunov spectra 
in one and two dimensional models satisfying  the total momentum 
conservation with periodic boundary conditions. 
   These models show a stepwise structure of the Lyapunov spectrum in the 
region of small positive Lyapunov exponents, which is robust to a perturbation 
in matrix elements of the particle interaction matrix. 
   The long range interactions between the particles lead to a clear separation 
into a part exhibiting stepwise structure and a part changing smoothly. 
   The part of the Lyapunov spectrum containing stepwise structure is 
clearly distinguished by a wave like structure in the 
eigenstates of the particle interaction matrix.
   In the two dimensional model we got wider steps in the Lyapunov spectrum 
than in the one dimensional models, especially the steps consisting of 4 points 
and the steps consisting of 8 points. 
   These wide steps in the Lyapunov spectrum have already been 
shown numerically in a deterministic chaotic system consisting 
of many hard-core disks.     

  One of the important simplifications in this random matrix approach
is that in this approach we do not have to refer to the phase 
space dynamics in order to determine the tangent space dynamics any more.   
   In general, the matrix $R(t)$ appearing in Eq. (\ref{Hamil}) can depend  
on the phase space dynamics, so this kind of separation of the phase space 
dynamics and the tangent space dynamics does not be allowed 
in deterministic chaotic systems. 

   As emphasized in the discussion of the two dimensional model, 
a lattice picture is useful to make a model in the random matrix approach. 
   Concerning this point it may be interesting to note that some works 
for the Lyapunov spectra for many particle systems indicated a similarity 
between solid state phenomena and the behavior of the Lyapunov 
spectra for many particle systems in a fluid phase \cite{Pos88,Hoo87}. 

   We have regarded the random matrix dynamics in this paper 
as an imitation of the deterministic chaotic dynamics, and reproduced some 
characteristics of chaotic systems, especially 
the stepwise structure of the Lyapunov spectra. 
   However we must not forget that there are some differences 
between the chaotic dynamics and the random matrix dynamics in this paper. 
   For example, in the random matrix dynamics the movement of particles 
are not deterministic but stochastic, so that the zero Lyapunov 
exponents arising from the initial infinitesimal perturbation along the orbit 
in the deterministic chaos do not appear in the random matrix dynamics. 
   (Note that we got only the zero Lyapunov exponents corresponding to the 
total momentum conservation in the models discussed in this paper.) 
   We should also mention that in the random matrix approach there 
is an additional statistical average over the randomness of the 
interactions of particles, which does not exist in the deterministic dynamics. 
   This causes vagueness in the definition of the Lyapunov exponents. 
   In this paper we introduced the Lyapunov exponent as the time 
average of {\em the logarithm of the randomness-averaged time evolution} of 
a neighboring trajectory. 
   This definition allows us to get a simple connection between the 
Lyapunov exponents and the time correlation of the interaction matrix 
as shown in Eq. (\ref{Lyapu2}). 
   However this definition of the Lyapunov exponent is not proper 
to discuss the negative Lyapunov exponents and hence the pairing rule 
for the Lyapunov spectrum. 
   On the other hand, we could adopt the definition of the Lyapunov 
exponent as the time average of { \em the randomness-average of 
the logarithm of time evolution} of a neighboring trajectory. 
   This definition should be proper to discuss the negative Lyapunov exponents. 
   The comparison of these two definitions of the Lyapunov exponent is 
an unsettled problem. 

   One may regard the random matrix approach using the master equation 
in this paper as one of the few analytical approaches to calculate 
the Lyapunov exponents. 
   The other statistical and analytical approach for the Lyapunov exponents  
is the kinetic theoretical approach of Refs. \cite{Bei95,Bei98}. 
   In this approach the positive Lyapunov exponents are calculated using 
a Lorentz-Boltzmann equation, while the negative Lyapunov exponents are 
calculated using an "anti-Lorentz-Boltzmann equation" where the collision 
operator has the opposite sign to the ordinary Lorentz-Boltzmann equation. 
   However, so far this kinetic theoretical approach can only provide 
the maximum Lyapunov exponent and the Kolmogorov-Sinai entropy 
for dilute gases. 

   To improve the random matrix approach used in this paper it is essential 
to know the statistical information for the interaction matrix $R(t)$. 
   It has already been shown numerically that the time average of the matrix $R(t)$ 
is almost null in the system consisting of many hard-core disks \cite{Mor01}. 
   This result justifies the condition (\ref{WhiteNoise1}) in the case of $n=1$. 
   A similar investigation of the correlation of the matrix $R(t)$ in deterministic 
   many particle chaotic systems is one of the important future problems.    


 

  
\vspace{0.6cm}
\begin{center}
{\large \bf Acknowledgments}
\end{center}

\vspace{0.2cm}
   One of the authors (T.T) wishes to thank C. P. Dettmann for 
   helpful discussions.
 
\vspace{0.3cm}


\appendix

 \setcounter{section}{0} 
 \makeatletter 
    \@addtoreset{equation}{section} 
    \makeatother 
    \def\theequation{\Alph{section}.%
    \arabic{equation}} 

\section{Master Equation for the Tangent Space}
\label{Maste} 

In this appendix we derive the Fokker-Plank equation (\ref{Fokke}) 
for the tangent vector space. 
Using the Kramers-Moyal expansion the dynamics of the probability 
density $\rho(\delta\bfGamma,t)$ is given by 

\begin{eqnarray}
   \frac{\partial \rho(\delta\bfGamma,t) }{\partial t} 
   = && \sum_{n=1}^{\infty} \sum_{j_{1}=1}^{2N} \sum_{j_{2}=1}^{2N} 
   \cdots 
   \sum_{j_{n}=1}^{2N} (-1)^{n} \nonumber \\
   && \hspace{0.5cm} \times
   \frac{\partial^{n}\Xi_{j_{1}j_{2}\cdots j_{n}}^{(n)} 
   (\delta\bfGamma,t) 
\rho(\delta\bfGamma,t)}{\partial\delta\caGamma_{j_{1}} 
\partial\delta\caGamma_{j_{2}}\cdots\partial\delta\caGamma_{j_{n}}  
   }  
\label{Krame}\end{eqnarray}

\noindent where $\Xi_{j_{1}j_{2}\cdots 
j_{n}}^{(n)}(\delta\bfGamma,t)$ 
is defined by 

\begin{eqnarray}
    && \Xi_{j_{1}j_{2}\cdots j_{n}}^{(n)}(\delta\bfGamma,t) \nonumber \\
   && \hspace{0.5cm}
   \equiv
      \frac{1}{n!} \lim_{s\rightarrow 0} \frac{1}{s} \Bigl\langle    
      [\delta\caGamma_{j_{1}}(t+s)-\delta\caGamma_{j_{1}}(t)]       
      \nonumber \\
   && \hspace{1.3cm}
      \times [\delta\caGamma_{j_{2}}(t+s)-\delta\caGamma_{j_{2}}(t)] 
      \nonumber \\
   && \hspace{1.3cm}  \left.\cdots   
      [\delta\caGamma_{j_{n}}(t+s)-\delta\caGamma_{j_{n}}(t)]
   \Bigr\rangle\right|_{\delta\smallbfGamma (t)=\delta\smallbfGamma}  
\label{ColliX}\end{eqnarray}

\noindent and $\delta\caGamma_{j}(t)$ is the $j$-th component of the 
tangent vector $\delta\bfGamma(t)$  \cite{ris89}.

     It follows from Eqs. (\ref{MatriJ}), (\ref{TangeEquat2}), (\ref{MatriM}) and 
(\ref{Hamil}) that   

\begin{eqnarray}
    &&\delta\bfGamma (t+s)-\delta\bfGamma (t) \nonumber \\
   && \hspace{0.5cm}
    =\left\{\stackrel{\leftarrow}{T}  
   \exp{\left[J \int_{t}^{t+s}d\tau\;L(\tau)\right]}-1 
   \right\}\delta\bfGamma (t)
   \nonumber \\
   && \hspace{0.5cm}
    = \sum_{n=1}^{\infty} \int_{t}^{t+s}d\tau_{n}
    \int_{t}^{\tau_{n}}d\tau_{n-1}
    \cdots \int_{t}^{\tau_{2}}d\tau_{1} \; 
      \nonumber \\
   && \hspace{0.5cm}
     \hspace{1cm}\times
    JL(\tau_{n})JL(\tau_{n-1})\cdots JL(\tau_{1}) \delta\bfGamma (t) 
\label{TimeOrderExpa}\end{eqnarray}

\noindent and 

\begin{eqnarray}
      &&
      JL(\tau_{n})JL(\tau_{n-1})\cdots JL(\tau_{1})  \nonumber \\
   \nonumber \\
   && \hspace{0.2cm} = \left\{ 
      \begin{array}{l}
   \left(
      \begin{array}{cc} 
      \underline{0} & I/m \\
      R(\tau_{1}) & \underline{0} 
      \end{array}\right) 
      \hspace{0.5cm} \mbox{for} \;\; n=1
   \\
   \\ 
      \left(
      \begin{array}{cc}
      \Phi_{l}^{(1)} & \underline{0} \\
      \underline{0} & \Phi_{l}^{(2)}
      \end{array} \right) 
      \hspace{0.5cm} \mbox{for} \;\; n=2l, l=1,2,\cdots 
   \\
   \\
      \left(
      \begin{array}{cc} 
      \underline{0} & \Phi_{l}^{(2)} /m \\
      m\Phi_{l+1}^{(1)} & \underline{0} 
      \end{array}\right) 
      \hspace{0.2cm} \mbox{for} \;\; n=2l+1, l=1,2,\cdots. 
   \end{array} \right. \nonumber \\
   &&
\label{MaltiMatriJL}\end{eqnarray}

\noindent where $\Phi_{l}^{(j)}$, $j=1,2$ are defined by 

\begin{eqnarray}
   \Phi_{l}^{(1)} \equiv  R(\tau_{2l-1})R(\tau_{2l-3})\cdots R(\tau_{1})
   /m^{l}
\label{FunctPhi1}\end{eqnarray}
\begin{eqnarray}
   \Phi_{l}^{(2)} \equiv R(\tau_{2l})R(\tau_{2l-2})\cdots R(\tau_{2})
   /m^{l}.
\label{FunctPhi2}\end{eqnarray}

\noindent By using Eqs. (\ref{WhiteNoise1}), 
(\ref{WhiteNoise2}), (\ref{ColliX}), (\ref{TimeOrderExpa})
 and  (\ref{MaltiMatriJL}) we obtain

\begin{eqnarray}
    &&\bfXi^{(1)}(\delta\bfGamma,t) \nonumber \\
   &&\hspace{0.2cm} \equiv (\Xi_{1}^{(1)}(\delta\bfGamma,t), 
      \Xi_{2}^{(1)}(\delta\bfGamma,t), 
	  \cdots,\Xi_{2N}^{(1)}(\delta\bfGamma,t) )^{T}
      \nonumber \\
   &&\hspace{0.2cm} =  
      \lim_{s\rightarrow 0} \frac{1}{s}
	  \left. \bigl\langle \left[\delta\bfGamma(t+s)-\delta\bfGamma(t) 
	  \right] \bigr\rangle \right|_{\delta\smallbfGamma(t)
	  =\delta\smallbfGamma}
	  \nonumber \\
   &&\hspace{0.2cm} =
      \lim_{s\rightarrow 0} \frac{1}{s} \int_{t}^{t+s} d\tau 
	  \left\langle J L(\tau) \right\rangle \delta\bfGamma
	  \nonumber \\
   &&\hspace{0.2cm} = (\delta p_{1}, \delta p_{2}
      \cdots,\delta p_{N}, 0, 0, \cdots, 0)^{T} /m
\label{FunctX1}\end{eqnarray}


\begin{eqnarray}
   && \Xi^{(2)}(\delta\bfGamma,t) \equiv 
   (\;\Xi_{j k}^{(2)}(\delta\bfGamma,t) \;) \nonumber \\
   &&\hspace{0.2cm}= 
      \lim_{s\rightarrow 0} \frac{1}{2s}
	  \biggl\langle 
	     \left[\delta\bfGamma(t+s)-\delta\bfGamma(t) \right] 
	  \nonumber \\
   &&\hspace{2cm} \times 
	   \left. \left[\delta\bfGamma(t+s)-\delta\bfGamma(t) \right]^{T} 
	  \biggr\rangle \right|_{\delta\smallbfGamma(t)
	     =\delta\smallbfGamma}
	  \nonumber \\
   &&\hspace{0.2cm} =
      \lim_{s\rightarrow 0} \frac{1}{2s} 
	  \int_{t}^{t+s} d\kappa \int_{t}^{t+s} d\tau 
	  \nonumber \\
   &&\hspace{2cm} \times 
	  \left\langle J L(\kappa) \delta\bfGamma 
	  \delta\bfGamma^{T} [J L(\tau)]^{T} \right\rangle 
	  \nonumber \\
   &&\hspace{0.2cm} = \left(\begin{array}{cc} 
      \underline{0} & \underline{0} \\
	  \underline{0} & \Psi^{(2)}(\delta\bfq)
    \end{array} \right)
\label{FunctX2}\end{eqnarray}

\noindent where $\Psi^{(2)}(\delta\bfq)\equiv 
(\Psi_{jk}^{(2)}(\delta\bfq))$ is defined by 

\begin{eqnarray}
   \Psi_{jk}^{(2)}(\delta\bfq)
   \equiv \frac{1}{2}
   \sum_{\alpha=1}^{N} \sum_{\beta=1}^{N} 
	     D_{j\alpha\beta k}\delta q_{\alpha} \delta q_{\beta}.
\label{MatriPsi}\end{eqnarray}


\noindent 
Here the only non zero contributions come from the $n=1$ term of Eq. 
(\ref{TimeOrderExpa}). For general $n$, the number of delta functions
from Eq. (\ref{WhiteNoise2}) must be only one less than the number 
of time integrals, to give a non
zero contribution. It is straightforward to show that this never 
happens for $n>1$.
   The terms including  $\Xi_{j_{1}j_{2}\cdots 
   j_{n}}^{(n)}(\delta\bfGamma,t)$,  $n=3,4,\cdots$ in 
the right-hand side of Eq. (\ref{Krame}) are negligible because 
of the Gaussian white properties (\ref{WhiteNoise1}) and 
(\ref{WhiteNoise2}) of 
the random matrix $R(t)$. 
   Using this fact and Eqs. (\ref{DiffuConst2}), (\ref{Krame}), 
(\ref{FunctX1}) and (\ref{FunctX2}) 
we obtain the Fokker-Planck equation (\ref{Fokke}).


\section{Lyapunov Exponents in the Random Matrix Approach}
\label{LyapuRando} 
   
   In this appendix we derive the expression (\ref{Lyapu2}) for the 
Lyapunov exponents from the definition (\ref{Lyapu}).  
   Transforming Eqs. (\ref{ConneUpsil1}), (\ref{ConneUpsil2}) and 
(\ref{ConneUpsil3}) into the equations for the quantities 
$\tilde{\Upsilon}_{jk}^{(l)}(t)$ we obtain 

\begin{eqnarray}
   \frac{d \tilde{\Upsilon}_{jk}^{(1)}(t)}{dt} &=& \frac{2}{m} 
       \tilde{\Upsilon}_{jk}^{(2)}(t)  \label{ConneUpsil1Trans} \\
   \frac{d \tilde{\Upsilon}_{jk}^{(2)}(t)}{dt} &=& \frac{1}{m} 
      \tilde{\Upsilon}_{jk}^{(3)}(t) \label{ConneUpsil2Trans} 
\end{eqnarray}
\begin{eqnarray} 
   \frac{d \tilde{\Upsilon}_{jk}^{(3)}(t) }{dt} 
	     &=& \sum_{\alpha=1}^{N}\sum_{\beta=1}^{N} 
         \sum_{\mu=1}^{N}\sum_{\nu=1}^{N}
         T_{j  \alpha \beta  k} D_{\alpha\mu\beta\nu} 
         \Upsilon_{\mu\nu}^{(1)}(t) \nonumber \\
	     &=& \sum_{\alpha=1}^{N}\sum_{\beta=1}^{N} 
         \sum_{\mu=1}^{N}\sum_{\nu=1}^{N}
         T_{j  \alpha \beta  k} D_{\alpha\mu\nu\beta} 
		 \nonumber \\
	     && \times\sum_{\alpha'=1}^{N}\sum_{\beta'=1}^{N} 
         \sum_{\mu'=1}^{N}\sum_{\nu'=1}^{N}
         T_{\mu'\alpha'\beta'\nu'}T_{\nu'\nu\mu\mu'}
		 \Upsilon_{\alpha'\beta'}^{(1)}(t) \nonumber \\
	     &=& \Lambda_{j k}Ÿ\tilde{\Upsilon}_{jk}^{(1)}(t) 
\label{ConneUpsil3Trans}\end{eqnarray}

\noindent where we used Eqs. (\ref{DiffuConst2}), (\ref{OrthoT1}), 
(\ref{OrthoT2}) and (\ref{TransUpsil}). 
   Equations (\ref{ConneUpsil1Trans}), (\ref{ConneUpsil2Trans}) 
and (\ref{ConneUpsil3Trans}) lead to 

\begin{eqnarray} 
    \frac{d^{3} \tilde{\Upsilon}_{jj}^{(1)}(t)}{dt^{3}} 
	&=& \frac{2\Lambda_{jj}}{m^{2}}   
	   Ÿ\tilde{\Upsilon}_{jj}^{(1)}(t)
       \label{ConneUpsil1a} \\
   \frac{d^{2}\tilde{\Upsilon}_{jj}^{(1)}(t)}{dt^{2}} &=& 
      \frac{2}{m^{2}} Ÿ\tilde{\Upsilon}_{jj}^{(3)}(t) 
       \label{ConneUpsil2a} \\
   \frac{d\tilde{\Upsilon}_{jj}^{(1)}(t) }{dt} 
      &=& \frac{2}{m} Ÿ\tilde{\Upsilon}_{jj}^{(2)}(t).
\label{ConneUpsil3a}\end{eqnarray}

\noindent It is noted that Eq. (\ref{ConneUpsil1a}) is  
the differential equation only for the quantity $\tilde{\Upsilon}_{jj}^{(1)}(t)$.

   If the quantity $\Lambda_{jj}$ is zero, then the quantity 
$\tilde{\Upsilon}_{jj}^{(1)}(t)$ is a bilinear function of 
time $t$ so that the Lyapunov exponent defined by Eq. 
(\ref{Lyapu}) gives zero, namely Eq.  (\ref{Lyapu2}) is correct in 
this case. 
   In the case of $\Lambda_{jj}\neq0$, noting that the functions 
$\exp[(2\Lambda_{jj}/m^{2})^{1/3} t \cdot \exp(2\pi k i/3)]$, 
$k=0,1,2$ are special solutions of Eq.  (\ref{ConneUpsil1a}), 
we obtain the general solution 

\begin{eqnarray} 
    \tilde{\Upsilon}_{jj}^{(1)}(t) 
    &=& \Omega_{j}^{(1)} \, \exp\left({\cal K}_{j} t \right) \nonumber \\
       && + \mbox{Re}\left\{\left(\Omega_{j}^{(2)} - i \Omega_{j}^{(3)} \right) 
          \exp\left[{\cal K}_{j} t \cdot \exp\left(\frac{2}{3}\pi i\right)
          \right]\right\}
          \nonumber  \\
    &=& \Omega_{j}^{(1)} \, \exp\left({\cal K}_{j} t \right) \nonumber \\
       && + \left[ \Omega_{j}^{(2)} \,  
          \cos \left(\frac{\sqrt{3}}{2} {\cal K}_{j}  t 
        \right) + \Omega_{j}^{(3)} \, 
             \sin \left(\frac{\sqrt{3}}{2} {\cal K}_{j} t \right)
          \right] \nonumber \\
       && \hspace{1cm} \times \exp\left(-\frac{1}{2}{\cal K}_{j} t \right) 
\label{Upsil(1)}\end{eqnarray}

\noindent of Eq.  (\ref{ConneUpsil1a}) 
for the real function $\tilde{\Upsilon}_{jk}^{(1)}(t)$ where  
${\cal K}_{j}$ is defined by 

\begin{eqnarray}
   {\cal K}_{j} \equiv  \left(\frac{2
           \Lambda_{jj}}{m^{2}} \right)^{1/3}.
\label{ConstK}\end{eqnarray}

\noindent Here $\Omega_{j}^{(k)}$, $k=1,2,3$ are real constants and 
are connected to the initial condition as 

\begin{eqnarray}
    \Omega_{j}^{(1)} &=& \frac{1}{3} \left[\tilde{\Upsilon}_{jj}^{(1)}(0) 
       + \frac{2\tilde{\Upsilon}_{jj}^{(2)}(0) }{m{\cal K}_{j}} 
	   + \frac{2\tilde{\Upsilon}_{jj}^{(3)}(0)}{(m{\cal K}_{j} )^{2}} 
            \right]
       \\
    \Omega_{j}^{(2)} &=& \frac{2}{3} \left[\tilde{\Upsilon}_{jj}^{(1)}(0) 
       - \frac{ \tilde{\Upsilon}_{jj}^{(2)}(0)}{m{\cal K}_{j} } 
	  - \frac{\tilde{\Upsilon}_{jj}^{(3)}(0)}{(m{\cal K}_{j} )^{2}} 
            \right]
       \\
   \Omega_{j}^{(3)} &=& \frac{2}{m{\cal K}_{j} \sqrt{3} } 
       \left[ \tilde{\Upsilon}_{jj}^{(2)}(0) 
		-\frac{\tilde{\Upsilon}_{jj}^{(3)}(0)}{m{\cal K}_{j} } 
            \right]
\label{InitiConsi}\end{eqnarray}

\noindent by using Eqs. (\ref{ConneUpsil2a}), (\ref{ConneUpsil3a}) and  
(\ref{Upsil(1)}).  
   By substituting Eq. (\ref{Upsil(1)}) into Eq. (\ref{Lyapu}) 
we obtain 

\begin{eqnarray}
   \lambda_{j} = \frac{{\cal K}_{j}}{2} 
   = \left[Ÿ\frac{\Lambda_{jj}}{(2m)^{2}} \right]^{1/3}, 
\label{Lyapu2Obtai}\end{eqnarray}

\noindent namely Eq. (\ref{Lyapu2}).

   Equations (\ref{ConneUpsil2a}) and (\ref{ConneUpsil3a}) imply that 
the quantities $\tilde{\Upsilon}_{jj}^{(2)}(t)$ and 
$\tilde{\Upsilon}_{jj}^{(3)}(t)$  are 
given by taking the first and second derivative of the quantity 
$\tilde{\Upsilon}_{jj}^{(1)}(t)$ with respect to the time $t$, 
respectively. 
   This fact leads to the relation 
  
\begin{eqnarray}
  \lim_{t\rightarrow\infty} 
   \frac{1}{2t} \ln \frac{\tilde{\Upsilon}_{jj}^{(1)}(t)}
   {\tilde{\Upsilon}_{jj}^{(1)}(0)}    
   &=& \lim_{t\rightarrow\infty} 
   \frac{1}{2t} \ln \frac{\tilde{\Upsilon}_{jj}^{(2)}(t)}
   {\tilde{\Upsilon}_{jj}^{(2)}(0)} \\
   &=& \lim_{t\rightarrow\infty} 
   \frac{1}{2t} \ln \frac{\tilde{\Upsilon}_{jj}^{(3)}(t)}
   {\tilde{\Upsilon}_{jj}^{(3)}(0)} 
\label{RelationUpsil(k)}\end{eqnarray}   
   
\noindent  Namely we get the same Lyapunov exponents as in Eq. 
(\ref{Lyapu})
through the equations $\lim_{t\rightarrow+\infty}$ $(2t)^{-1}\ln 
   [\tilde{\Upsilon}_{jj}^{(k)}(t) ) /
   \tilde{\Upsilon}_{jj}^{(k)}(0) ] $, $k=2,3$.


\section{Lyapunov Exponents for One Dimensional Models 
with the Nearest Neighbor Interactions}
\label{LyapuOneDimen} 
   
   In this appendix we give the derivation of Eqs. (\ref{LyapuModel1}) 
and (\ref{LyapuDegen})
from Eqs. (\ref{Lyapu3}) and (\ref{MatriA0}). 
   We also show Eqs. (\ref{FirstOrderLyapu1}) and (\ref{FirstOrderLyapu2})
in the case of $A=A^{(0)}+\varepsilon A^{(1)}$. 

   We define the discretized Fourier-transform matrix 
$F\equiv (F_{jk})$ by 

\begin{eqnarray}
   F_{kl} \equiv \frac{1}{\sqrt{N}} \exp\left(-2\pi i \, 
   \frac{kl}{N}\right), 
\label{FouriMatri}\end{eqnarray} 

\noindent which is an unitary matrix: $F^{\dagger} F = F 
F^{\dagger} = I$ with the superscript $\dagger$ representing 
the Hermitian conjugate of the matrix.

   First, it follows from Eqs. (\ref{MatriA0}) and (\ref{FouriMatri}) that 

\begin{eqnarray}
    && \sum_{\alpha=1}^{N} A^{(0)}_{j\alpha} F_{\alpha k}  
   \nonumber \\
   & & \hspace{0.5cm} = \frac{\omega}{\sqrt{N}} 
      \bigl[-2 \exp(-2\pi i jk/N)    \nonumber \\
   & & \hspace{1cm} \left.  
      +\theta_{Nj}\theta_{j2} \; \exp(-2\pi i (j-1) k/N) 
      \right. \nonumber \\
   & & \hspace{1cm} \left. 
      +\theta_{(N-1)j}\theta_{j1} \; \exp(-2\pi i (j+1) k/N) 
      \right. \nonumber \\
   & & \hspace{1cm}  
      \left.  +\theta_{Nj}\theta_{jN} \; \exp(-2\pi i (j-N+1) k/N)
        \right. \nonumber \\
   & & \hspace{1cm} 
      +\theta_{1j}\theta_{j1} \; \exp(-2\pi i (j+N-1) k/N) \bigr] 
      \nonumber \\
   & & \hspace{0.5cm}  = a_{k}^{(0)} F_{jk}  
\label{DiagoA0}\end{eqnarray}
 
\noindent where $\theta_{jk}$ is defined by 

\begin{eqnarray}
   \theta_{jk} \equiv \left\{
   \begin{array}{ll}
      1 & \mbox{in}\; j\geq k \\
      0 & \mbox{in}\; j<k 
   \end{array}
   \right.   
\label{FunctTheta}\end{eqnarray}
 
\noindent and satisfies the relations 
$\theta_{jk}\theta_{kj}=\delta_{jk}$ and 
$\theta_{N n}\theta_{n l} + \theta_{(l-1) n}\theta_{n1} = 1$ for 
any integer $l$$\in\{2,3,\cdots,N]$ and any integer 
$n$$\in\{1,2,\cdots,N\}$.  
   Here the $k$-th eigenvalue $a_{k}^{(0)}$ of the matrix $A^{(0)}$ is 
given by 

\begin{eqnarray}
   a_{k}^{(0)} = -2 \omega \left( 1-\cos\frac{2\pi k}{N} \right).
\label{EigenOneDimen1}\end{eqnarray}

\noindent  
   By using the formula (\ref{Lyapu3}) for the eigenvalues 
$a_{j}=a_{j}^{(0)}$ given by Eq. (\ref{EigenOneDimen1})
we obtain Eqs. (\ref{LyapuModel1}).
   The eigenvalues of the matrix $A^{(0)}$ 
have degeneracies because of the 
relation  $a_{N-j}^{(0)} = 
a_{j}^{(0)}$ in $j<N/2$, so we obtain Eq. (\ref{LyapuDegen}).

   Equation (\ref{DiagoA0}) implies that the matrix $A^{(0)}$ is diagonalized 
by using the matrix $F$: 
$(F^{\dagger} A^{(0)} F)_{jk}Ÿ = a_{j}^{(0)}Ÿ \delta_{jk}$. 
   The eigenvectors $\bfx_{j}^{(0)}$ of the 
matrix $A^{(0)}$ corresponding to the eigenvalue $a_{j}^{(0)}$
is represented as 

\begin{eqnarray}
   \bfx_{j}^{(0)} = (F_{1j}, F_{2j}, \cdots, F_{Nj})^{T},  
 \label{EigenVectorA0}\end{eqnarray} 
   
\noindent so that we have 
the relation $A^{(0)} \bfx_{n}^{(0)} = a_{j}^{(0)} \bfx_{n}^{(0)}$.
  This set of eigenvectors satisfies the completeness condition  
and is normalized, namely $\sum_{\alpha=1}^{N} \bfx_{\alpha}^{(0)} 
\bfx_{\alpha}^{(0)} { }^{\dagger}  = I$ and 
$\bfx_{j}^{(0)} { }^{\dagger} \bfx_{k}^{(0)} =\delta_{jk}$.
  
   Second, we expand the $j$-th eigenvalue $a_{j}$  of the matrix 
$A=A^{(0)} + \varepsilon A^{(1)}$  in the small parameter 
$\varepsilon$, namely $a_{j} = a_{j}^{(0)} +\varepsilon 
a_{j}^{(1)} + \cdots$, and consider the first order correction 
$\varepsilon a_{j}^{(1)}$ by using the well-known perturbation theory 
for a degenerate system. 
   We introduce the eigenstate $\bfx_{j}$ of the matrix $A$ 
corresponding 
to the eigenvalue $a_{j}$, and expands it with the complete set  
$\{\bfx_{l}^{(0)}\}_{l}$ of the vectors: 

\begin{eqnarray}
\bfx_{j} = \sum_{\alpha=1}^{N} 
c_{j\alpha} \bfx_{\alpha}^{(0)}
\label{EigenStateExpan}\end{eqnarray}

\noindent  with the constant $c_{j\alpha} 
\equiv \bfx_{\alpha}^{(0) \dagger}\bfx_{j}$. 
   The relation $A\bfx_{j}=a_{j}\bfx_{j}$ is translated into 
the equation 
   
\begin{eqnarray}
    \sum_{\alpha=1}^{N}\left[(a_{k}^{(0)} - a_{j})\delta_{k\alpha} +\varepsilon 
    W_{k\alpha} \right] c_{j\alpha} =0
\label{EigenEquati}\end{eqnarray}

\noindent for the coefficients $\{c_{jk}\}_{j,k}$ and the eigenvalue $a_{j}$ 
with the quantity 

\begin{eqnarray}
W_{jk}\equiv\bfx_{j}^{(0)}{ }^{\dagger}ŸA^{(1)} 
\bfx_{k}^{(0)}.
\label{QuantW}\end{eqnarray}

\noindent We expand the coefficient $c_{jk}$ by the small parameter 
$\varepsilon$: $c_{jk} = c_{jk}^{(0)} + \varepsilon \, c_{jk}^{(1)} 
+\cdots$. 
   Now we calculate the first order corrections 
$\varepsilon a_{j}^{(1)}$ and $\varepsilon a_{N-j}^{(1)}$ of the 
eigenvalues $a_{j}$ and $a_{N-j}$, respectively,  which
have a degeneracy in the 0-th order of the parameter 
$\varepsilon$. 
   For this purpose,  instead of the coefficients
$\{c_{jk} \}_{j,k}$ appearing in Eq. (\ref{EigenEquati}) 
it is enough to consider the 0-th order coefficients
$\{c_{jk}^{(0)}\}_{j,k}$, which are zero except for 
$c_{jj}^{(0)}$, $c_{j(N-j)}^{(0)}$, $c_{(N-j)j}^{(0)}$ and 
$c_{(N-j) (N-j)}^{(0)}$. 
 This leads to the equation  

\begin{eqnarray}
   \mbox{Det}\left( \begin{array}{cc}
       - a^{(1)} + W_{j j} &  W_{j (N-j)} \\ 
        W_{(N-j) j} & - a^{(1)} + W_{(N-j) (N-j)}  
   \end{array} \right) = 0 
\label{EigenEquati2}\end{eqnarray}

\noindent for $a^{(1)}$, whose solutions give the eigenvalues 
$a_{j}^{(1)}$ and  $a_{N-j}^{(1)}$.  
   We can solve Eq. (\ref{EigenEquati2}) and obtain
   
\begin{eqnarray}
   a^{(1)} &=&  \frac{W_{j j} + W_{(N-j)(N- j)}}{2} \nonumber \\ 
   && \pm 
   \sqrt{\left( \frac{W_{j j} - W_{(N-j)(N- j)}}{2}\right)^{2} 
   +\left|W_{j(N-j)}\right|^{2}}, \nonumber \\ &&
\label{SolutEquat}\end{eqnarray}

\noindent  noting the relation $W_{kj}=W_{jk}^{*}$ 
with the superscript $*$ representing the complex conjugate 
of the complex number.
   The eigenvalue $a_{N}^{(0)}$ has no degeneracy so we 
simply have 
$a_{N}^{(1)} = W_{NN}$.
   If the number N is even, then the eigenvalue $a_{N/2}^{(0)}$ has 
no degeneracy, so we also have $a_{N/2}^{(1)} =  
W_{(N/2)(N/2)}$ in this case.
  
   By using Eqs. (\ref{MatriA1}),   (\ref{FouriMatri}), 
(\ref{EigenVectorA0}) and (\ref{QuantW}) we obtain 
    
\begin{eqnarray}
   W_{kl} 
   &=&  - \left[ 1 - 
   \exp\left(- 2\pi i k/N \right) \right] \nonumber \\
   &&   \hspace{0cm} \times \;
   \left[ 1 - \exp\left(2\pi i l/N \right) \right] N^{-1} 
   \nonumber \\
   &&   \hspace{0cm} \times \sum_{\alpha=1}^{N} 
   \chi_{\alpha} \; \exp\left[2\pi i 
\alpha(k-l) /N \right]. 
\label{FouriMatriA1} \end{eqnarray} 
    
\noindent Especially we derive 

\begin{eqnarray}
   W_{kk} 
   &=& - \left| 1 - \exp\left(- 2\pi i k/N \right) \right|^{2} 
   \tilde{\chi}_{0} \nonumber \\
   &=& \frac{\tilde{\chi}_{0}}{\omega} a_{k}^{(0)} =  W_{(N- k)(N-k)} 
\label{matriW1} \end{eqnarray}
\begin{eqnarray}
   W_{k(N-k)} 
      = - \left[1-\exp\left(-2\pi i k/N \right)\right]^{2} 
      \tilde{\chi}_{k}
\label{matriW2}\end{eqnarray}

\noindent from Eqs. (\ref{TildeChi}), (\ref{EigenOneDimen1}) 
and (\ref{FouriMatriA1}).
   By substituting Eqs. (\ref{matriW1}) and (\ref{matriW2}) into 
Eq. (\ref{SolutEquat}) we obtain 

\begin{eqnarray} 
   a_{j}^{(1)} = \frac{ a_{j}^{(0)}}{\omega} 
    \left(\tilde{\chi}_{0} + |\tilde{\chi}_{j}|\; \right) 
    \label{FirstOrder1}  \\
   a_{N-j}^{(1)} = \frac{ a_{N-j}^{(0)}}{\omega}
   \left(\tilde{\chi}_{0} - |\tilde{\chi}_{j}|\; \right)
\label{FirstOrder2} \end{eqnarray} 

\noindent in the case of $j<N/2$, where we used 
the relation $|W_{j(N-j)}| = 
|\tilde{\chi}_{j} a_{j}^{(0)} /\omega|$ and 
numbered so that we obtain  
$(a_{j}^{(1)} - a_{N-j}^{(1)}) \omega /a_{j}^{(0)}\geq 0$.
   By using Eq. (\ref{matriW1}) we also obtain  $a_{N/2}^{(1)} =  
\tilde{\chi}_{0} a_{N/2}^{(0)}/\omega$
in the case of the number N to be even, and  $a_{N}^{(1)}=0$ for the 
first order correction to the non-degenerate eigenvalues of the 
matrix $A^{(0)}$. 
  By applying the formula (\ref{Lyapu3}) to the case of $a_{j} 
=a_{j}^{(0)} +\varepsilon a_{j}^{(1)} + {\cal O}(\varepsilon^2$) 
with the quantity $a_{j}^{(1)}$ given by 
Eq. (\ref{FirstOrder1}) or (\ref{FirstOrder2}), we obtain 
Eqs. (\ref{FirstOrderLyapu1}) and  (\ref{FirstOrderLyapu2}).


\section{Lyapunov Exponents for One Dimensional Models 
 with Long Range Interactions of a Fixed Strength}
\label{LyapuOneDimenLong} 

   In this appendix we calculate the Lyapunov spectra 
for the systems with long range interactions of a fixed  
strength. 

First we consider the case described by the matrix  
$A=\bar{A}^{[n_{l}]}$ $\equiv (\bar{A}^{[n_{l}]}_{jk})$ 
defined by 

\begin{eqnarray}
    \bar{A}_{jk}^{[n_{l}]} &=&  
    -2n_{l} \bar{\omega} \delta_{jk} 
    + \bar{\omega} \sum_{l=1}^{n_{l}} \left[
       \delta_{j(k+l)} + \delta_{k(j+l)} \right. \nonumber \\
	   &&\left.
       + \delta_{j(k+N-l)} + \delta_{k(j+N-l)}  
    \right] 
\label{MatriA2long}\end{eqnarray} 

\noindent with $n_{l}<N/2$ and  a (non-zero) real constant 
$\bar{\omega}$. 
   It should be noted that the matrix $A^{[n_{l}]}$ given by 
Eq. (\ref{MatriA2}) is attributed into this matrix $\bar{A}^{[n_{l}]}$ 
in the case of $\sigma_{j}^{[l]} =  \bar{\omega}$.
   The matrix $\bar{A}^{[n_{l}]}$ is diagonalized by 
the matrix $F$ defined by Eq. (\ref{FouriMatri}), and  
we obtain the relation  
$(F^{\dagger}\bar{A}^{[n_{l}]}F)_{jk} =\bar{a}_k^{[n_{l}]} 
\delta_{jk}$ with the eigenvalue 
$
   \bar{a}_{j}^{[n_{l}]} \equiv -2\bar{\omega} 
   \left[
      n_{l} - \sum_{l=1}^{n_{l}} 
      \cos\left(2\pi jl/N\right)
   \right]. 
$
\noindent 
By applying the formula (\ref{Lyapu3}) to 
the eigenvalue $a_{j}= \bar{a}_{j}^{[n_{l}]}$ 
we obtain the Lyapunov exponent  

\begin{eqnarray}
    \lambda_{j} 
   = \left|\frac{\bar{\omega}}{m}\right|^{2/3} 
   \left[ n_{l} - \sum_{l=1}^{n_{l}} \cos\left(\frac{2\pi jl}{N}\right)
   \right]^{2/3}.
\label{MatriA2long2}\end{eqnarray} 

\noindent The Lyapunov exponents given by Eq. (\ref{MatriA2long2}) satisfies 
the relation $\lambda_{j}=\lambda_{N-j}$ in $j<N/2$, so the 
Lyapunov spectrum of this system has a stepwise structure.

   Second we consider the case that 
each particle interacts with all the other particles with the same strength.  
   This is described by the matrix 
$A=\bar{A}^{[N/2]}$ $\equiv (\bar{A}^{[N/2]}_{jk})$, 
which is defined by 

\begin{eqnarray}
   \bar{A}^{[N/2]}_{jk} \equiv 
   2 \bar{\omega} (1-N \delta_{jk}), 
\label{MatriALimit}\end{eqnarray}

\noindent namely the matrix whose off-diagonal elements 
are non-zero and equal with each other.  
   The matrix (\ref{MatriALimit}) is diagonalized as 

\begin{eqnarray}
   (F^{\dagger} \bar{A}^{[N/2]}F)_{jk}
   = -2 N\bar{\omega}  (1-\delta_{jN}) \delta_{jk}
\label{MatriALimitEigen}\end{eqnarray}

\noindent by using the matrix $F$ defined by Eq. (\ref{FouriMatri}), 
so  the eigenvalues of the matrix $\bar{A}^{[N/2]}$ are 
$-2N\bar{\omega}$ and $0$.
 By substituting the eigenvalues of the matrix $A=A^{[N/2]}$ 
in the formula (\ref{Lyapu3}) we obtain the Lyapunov spectrum as 

\begin{eqnarray}
   \lambda^{[j]} = \left\{
   \begin{array}{cl}
      \left| \frac{N \bar{\omega}}{m} \right|^{2/3}  
      & \mbox{in} \; j=1,2, \cdots N-1 \\
      0  & \mbox{in} \; j = N ,
   \end{array}
   \right.
\label{MatriALimitLyapu}\end{eqnarray}

\noindent namely the shape of positive Lyapunov spectrum 
in this system is just in a straight horizontal line.     
   It should be noted that the quantity $\bar{\omega}$ 
may depend on the number $N$ of the particles in general. 
   Therefore the consideration in this appendix is not enough 
to discuss the particle number dependence of the maximum 
Lyapunov exponent.



\end{multicols}


\begin{thebibliography}{99}
\bibitem{note1} Equation (\ref{OrthoT2}) is corresponding to the  
 diagonalization of the matrix $\tilde{\Lambda} \equiv 
 \lim_{t\rightarrow +\infty} [M(t)^{T} 
 M(t)]^{1/t}$ in calculation of the Lyapunov exponents introduced as 
 the eigenvalues of the matrix $(\ln \tilde{\Lambda})/2$ 
 in a deterministic chaos.
\end{thebibliography}
\end{document}